\documentclass[12pt]{article}
\usepackage{latexsym,amsmath,amssymb,graphicx}
\renewcommand{\thefootnote}{\fnsymbol{footnote}}

\usepackage{amsmath}
\usepackage{amsfonts}

\numberwithin{equation}{section}

\def\doubleset#1#2{\bgroup%
\def\doit#1#2{%
\setbox\dblsetbox=\hbox{$\cstyle #1$}%
\raise#2\ht\dblsetbox\copy\dblsetbox%
\hskip-\wd\dblsetbox%
\raise-#2\ht\dblsetbox\box\dblsetbox}%
\mathchoice%
{\def\cstyle{\displaystyle}\doit#1#2}%
{\def\cstyle{\textstyle}\doit#1#2}%
{\def\cstyle{\scriptstyle}\doit#1#2}%
{\def\cstyle{\scriptscriptstyle}\doit#1#2}\egroup}
\def\underarrow#1{\vbox{\ialign{##\crcr$\hfil\displaystyle
 {#1}\hfil$\crcr\noalign{\kern1pt\nointerlineskip}$\longrightarrow$\crcr}}}

\newbox\dblsetbox

\newlength{\extraspace}
\newlength{\extraspaces}
\newcommand{\be}{\begin{equation}
\addtolength{\abovedisplayskip}{\extraspaces}
\addtolength{\belowdisplayskip}{\extraspaces}
\addtolength{\abovedisplayshortskip}{\extraspace}
\addtolength{\belowdisplayshortskip}{\extraspace}}
\newcommand{\ee}{\end{equation}}

\newcommand{\ba}{\begin{eqnarray}
\addtolength{\abovedisplayskip}{\extraspaces}
\addtolength{\belowdisplayskip}{\extraspaces}
\addtolength{\abovedisplayshortskip}{\extraspace}
\addtolength{\belowdisplayshortskip}{\extraspace}}
\newcommand{\ea}{\end{eqnarray}}

\newcommand{\bd}{\begin{displaymath}
\addtolength{\abovedisplayskip}{\extraspaces}
\addtolength{\belowdisplayskip}{\extraspaces}
\addtolength{\abovedisplayshortskip}{\extraspace}
\addtolength{\belowdisplayshortskip}{\extraspace}}
\newcommand{\ed}{\end{displaymath}}

\newcounter{saveeqn}

\newcommand{\newsection}[1]{
\vspace{12mm} \pagebreak[3] \addtocounter{section}{1}
\setcounter{equation}{0} \setcounter{subsection}{0}
\noindent{\bf \thesection. #1} \nopagebreak
\medskip
\nopagebreak}

\newcommand{\newsubsection}[1]{
\vspace{0.8cm} \pagebreak[3] \addtocounter{subsection}{1}
\noindent{\it \thesubsection. #1} \nopagebreak \vspace{2mm}
\nopagebreak}

\begin{document}
\addtolength{\baselineskip}{1.5mm}

\thispagestyle{empty}
%\begin{flushright}
%hep-th/   \\
%\end{flushright}
\vbox{} \vspace{1.5cm}

\begin{center}
\centerline{\LARGE{Gauging Spacetime Symmetries On The
Worldsheet}}
\bigskip
\centerline{\LARGE{And The Geometric Langlands Program -- II}}

\vspace{2.5cm}

{Meng-Chwan~Tan \footnote{On leave of absence from the National
University of Singapore.}}
\\[2mm]
{\it School of Natural Sciences, Institute for Advanced Study\\
Princeton, New Jersey 08540, USA} \\[1mm]
e-mail: tan@ias.edu\\
\end{center}

\vspace{2.5 cm}

\centerline{\bf Abstract}\smallskip \noindent

We generalise the analysis carried out in \cite{Langlands 1}, and find that our previous results can be extended beyond the case of $SL(N, \bf{C})$. In particular, we show that an equivalence---at the level of the holomorphic chiral algebra---between a bosonic string on a smooth coset manifold $G/B$ and  a $B$-gauged version of itself on $G$, will imply an isomorphism of classical $\cal W$-algebras and a level relation which underlie a geometric Langlands correspondence for the simply-laced, complex $ADE$-groups. In addition, as opposed to line operators and branes of an open topological sigma-model, the Hecke operators and Hecke eigensheaves, can, instead, be physically interpreted in terms of  the correlation functions of local operators in the holomorphic chiral algebra of a {\it{closed}}, {\it{quasi-topological}}  sigma-model. Our present results thus serve as an $\it{alternative}$ physical interpretation---to that of an electric-magnetic duality of four-dimensional gauge theory demonstrated earlier by Kapustin and Witten in \cite{KW}---of the geometric Langlands correspondence for complex $ADE$-groups. The cases with tame and mild ``ramifications'' are also discussed.

\newpage

\renewcommand{\thefootnote}{\arabic{footnote}}
\setcounter{footnote}{0}

\newsection{Introduction}

The geometric Langlands correspondence has recently been given an
elegant physical interpretation by Kapustin and Witten in their
seminal paper \cite{KW}---by considering a certain twisted ${\cal
N}=4$ supersymmetric Yang-Mills theory in four-dimensions
compactified on a complex Riemann surface $C$, the geometric
Langlands correspondence associated to a holomorphic $G$-bundle on
$C$ can be shown to arise naturally from an electric-magnetic
duality in four-dimensions. Specifically, it was first argued that one can, among other things, relate various mathematical
objects and concepts of the correspondence such as Hecke
eigensheaves and the action of the Hecke operator, to the
boundaries and the 't Hooft line operator of the underlying
four-dimensional quantum gauge theory. It was then shown that the map between the various ingredients which defines the mathematical correspondence, is nothing but a four-dimensional
electric-magnetic duality, or equivalently, a mirror symmetry of the resulting
two-dimensional topological sigma-model at low-energies. The framework outlined in \cite{KW} thus furnishes a purely physical interpretation of the geometric Langlands conjecture.

The work of Kapustin and Witten centres around a gauge-theoretic
interpretation of the geometric Langlands correspondence. However,
it does not shed any light on the utility of two-dimensional
$\it{axiomatic}$ conformal field theory in the geometric Langlands
program, which, incidentally, is ubiquitous in the mathematical
literature on the subject \cite{Beilinson, 2a, 2b, 2c, 2d, Frenkel}. This
seems rather puzzling. Afterall, the various axiomatic definitions
of a conformal field theory that fill the mathematical literature,
are based on established physical concepts, and it is therefore
natural to expect that in any physical interpretation of the
geometric Langlands correspondence, a two-dimensional conformal
field theory of some sort will be involved. It will certainly be
illuminating for the geometric Langlands program as a whole, if
one can deduce the axiomatic conformal field-theoretic approach developed in
the mathematical literature, from the gauge-theoretic approach of
Kapustin and Witten, or vice-versa.

In the axiomatic conformal field-theoretic approach to
the geometric Langlands correspondence, the key ingredients are affine Lie algebras at
the critical level without stress tensors \cite{book}, and $\cal
W$-algebras (defined by a Drinfeld-Sokolov or DS reduction
procedure) associated to the affine versions of the Langlands dual
of the Lie algebras \cite{book, dikii}. The duality between
$\it{classical}$ $\cal W$-algebras---which underlies the axiomatic conformal
field-theoretic approach to the correspondence---is just an
isomorphism between the $\it{Poisson}$ algebra generated by the
centre ${\frak z}(\widehat {\frak g})$ of the $\textrm{\it
completed universal enveloping algebra}$ of the affine Lie algebra
$\widehat {\frak g}$ at the $\it{critical}$ level, where $\frak g$ is the
Lie algebra of the group $G$, and the classical $\cal
W$-algebra associated to the affine Lie algebra $^L {\widehat
{\frak g}}$ in the limit of large level $k'$ -- ${\cal
W}_{\infty}(^L\widehat {\frak g})$, where $^L{\frak g}$ is the
Lie algebra of the Langlands dual group $^LG$; in other
words, a geometric Langlands correspondence for $G$ simply
originates from an isomorphism ${\frak z}(\widehat {\frak g})
\simeq {\cal W}_{\infty}(^L\widehat {\frak g})$ of $\it{Poisson}$
algebras \cite{Frenkel, Langlands-Drinfeld}. This statement is
accompanied by a relation $(k+h^{\vee})r^{\vee}= (k' +
{^Lh}^{\vee})^{-1}$ between the generic levels $k$ and $k'$ of
$\widehat {\frak g}$ and $^L{\widehat {\frak g}}$ respectively
(where $r^{\vee}$ is the lacing number of $\frak g$, and
$h^{\vee}$ and ${^Lh}^{\vee}$ are the dual Coxeter numbers of
$\frak g$ and $^L{\frak g}$).

Note that the gauge-theoretic approach to the program necessarily
involves a certain two-dimensional quantum field theory in its
formulation, a generalised topological sigma-model to be exact.
This strongly suggests that perhaps a good starting point towards
elucidating the connection between the axiomatic conformal field-theoretic
and gauge-theoretic approaches, would be to explore other physical
models which are $\it{purely}$ two-dimensional, that will enable us to make direct
contact with the central results of the correspondence derived
from the axiomatic conformal field-theoretic approach.

A strong hint that one should be considering for this purpose a
two-dimensional twisted $(0,2)$ sigma-model on a flag manifold,
stems from our recent understanding of the role sheaves of
``Chiral Differential Operators" (or CDO's) play in the
description of its holomorphic chiral algebra \cite{CDO}, and from
the fact that global sections of CDO's on a flag manifold furnish
a module of an affine Lie algebra at the critical level \cite{CDO,
MSV}. On the other hand, since Toda field theories lead to
free-field realisations of the $\cal W$-algebras defined by the DS
reduction scheme mentioned above (see Sect. 6 of \cite{review},
and the references therein), and since the Toda theory can be
obtained as a gauge-invariant content of a certain gauged WZW
theory \cite{review 23, review 24},  it should be true that a physical
manifestation of the isomorphism of (classical) $\cal W$-algebras which underlie the geometric Langlands correspondence, ought to be given by some relation between the sigma-model on a flag manifold and a gauged WZW model. This was the main motivation for the work in \cite{Langlands 1}, which represents a modest attempt towards an analysis of the relation between quantum field theory and the geometric Langlands correspondence from a purely two-dimensional viewpoint, wherein a twisted $(0,2)$ sigma-model on a complex flag manifold of $SL(N,\bf{C})$ was considered.

In this paper, we shall generalise the analysis in \cite{Langlands 1}, and show that our previous results can be extended beyond $SL(N, \bf{C})$ to include all complex simply-laced groups. In particular, we shall show that an equivalence---at the level of the $\it{holomorphic}$ chiral algebra---between a bosonic string on a smooth coset manifold $G/B$ and a $B$-gauged version of itself on $G$, will necessarily imply an isomorphism ${\frak z}(\widehat {\frak g})\simeq {\cal W}_{\infty}(^L\widehat {\frak g})$ of classical $\cal W$-algebras  and the relation $(k+h^{\vee})r^{\vee}= (k' + {^Lh}^{\vee})^{-1}$ which underlie a geometric Langlands correspondence for $G$, where $G$ is any simply-laced, complex $ADE$-group. This equivalence in the spectra of the bosonic strings---which can be viewed as a consequence of the ubiquitous notion that one can always physically interpret a geometrical symmetry of the target space as a gauge symmetry in the worldsheet theory---thus furnishes an $\it{alternative}$ physical interpretation, to that of an electric-magnetic duality of four-dimensional gauge theory, of the geometric Langlands correspondence for the complex $ADE$-groups! In addition, as in \cite{Langlands 1}, the Hecke operators and Hecke eigensheaves  of the geometric Langlands program for $G$, can also be shown to lend themselves to different physical interpretations altogether---instead of line operators and branes in a two-dimensional topological sigma-model, they are, in our context, related to the correlation functions of local operators that span the holomorphic chiral algebra of a $\it{closed}$ and {\it{quasi-topological}} sigma-model in two-dimensions. Moreover, the cases with tame and mild ``ramifications'' can also be understood from a purely physical perspective via these local operators. Our results therefore open up an alternative way of looking at the correspondence from a purely two-dimensional quantum field-theoretic standpoint, which could potentially lead to novel mathematical and physical insights for the geometric Langlands program as a whole.

\smallskip\noindent{\it A Brief Summary and Plan of the Paper}

We shall now give a brief summary and plan of the paper.

In $\S$2, we begin by considering the twisted $(0,2)$ sigma-model on a complex flag manifold given by the coset space $G/B$, where $G$ is any simply-laced, complex $ADE$-group with $\frak g = {^L{\frak g}}$, and $B$ is a Borel subgroup containing upper triangular matrices of $G$. We will show that the Casimir fields spanning the $\it{classical}$ holomorphic chiral algebra of the purely bosonic sector of the sigma-model, will have Laurent modes that generate the $\it{classical}$ centre $\frak{z}(\widehat{\frak g})$ of the completed universal enveloping algebra of the affine $G$-algebra at the critical level.

In $\S$3, we discuss the $\it{dual}$ description of the holomorphic chiral algebra of the purely bosonic
sector of the sigma-model on $G/B$, given by the $\it{holomorphic}$ BRST-cohomology (or chiral
algebra) of a $B$-gauged WZW model on $G$. We then show that the holomorphic BRST-cohomology of the $B$-gauged WZW model on $G$ at level $k'$ physically realises, in all generality, the Hecke algebra of local operators---generated by a mathematical Drinfeld-Sokolov reduction procedure \cite{book}---which defines ${\cal W}_{k'}(\widehat {\frak g})$, the $\cal W$-algebra associated to ${\widehat {\frak g}}$ at level $k'$.

In $\S$4, we use the results in the earlier sections to show that an equivalence---at the level of the holomorphic chiral algebra---between a bosonic string on $G/B$ and a $B$-gauged
version of itself on $G$, will necessarily imply an isomorphism ${\frak z}(\widehat {\frak g})
\simeq {\cal W}_{\infty}(^L\widehat {\frak g})$ of classical $\cal W$-algebras and the relation $(k+h^{\vee})r^{\vee}= (k' + {^Lh}^{\vee})^{-1}$ which underlie a
geometric Langlands correspondence for $G$.

In $\S$5, we shall derive, via the isomorphism ${\frak z}(\widehat {\frak g})
\simeq {\cal W}_{\infty}(^L\widehat {\frak g})$ of classical $\cal
W$-algebras, a correspondence between flat
holomorphic $^LG$-bundles on the worldsheet $\Sigma$ and Hecke
eigensheaves on the moduli space $\textrm{Bun}_G$ of holomorphic
$G$-bundles on $\Sigma$. Then, we shall
physically interpret the Hecke eigensheaves and Hecke operators of
the geometric Langlands program in terms of the correlation
functions of purely bosonic local operators in the holomorphic
chiral algebra of the twisted $(0,2)$ sigma-model on the complex
flag manifold $G/B$.

In $\S$6, we shall briefly discuss the physical interpretation of the geometric Langlands correspondence for $G$ with tame and mild ``ramifications'', in our setting.

%In $\S$6, we will conclude the paper with some remarks and open questions.

%In appendix~A, we will review the relevant characteristics of the two-dimensional twisted $(0,2)$
%sigma-model considered in \cite{CDO}, and explain its relation to
%the theory of CDO's. In particular, we will describe how its holomorphic chiral algebra can be interpreted in terms of %the sheaf of CDO's and its Cech-cohomology.

\smallskip\noindent{\it Relation to the Gauge-Theoretic Approach}

Though we have not made any explicit connections to the
gauge-theoretic approach of Kapustin and Witten yet, we hope to be
able to address this important issue in a later publication,
perhaps with the insights gained in this paper.

\newsection{The Twisted $(0,2)$ Sigma-Model on $G/B$ and the Classical Centre $\frak{z}(\widehat{\frak g})$}

In this section, we consider the twisted $(0,2)$ sigma-model on a complex flag manifold given by the coset space $G/B$, where $G$ is any complex $ADE$-group and $B$ is a Borel subgroup containing upper triangular matrices of $G$. Via a mathematical theorem in \cite{MSV}, and the interpretation of the $\overline Q_+$-cohomology of the sigma-model as the Cech-cohomology of the sheaf of CDO's (as reviewed in appendix~A of \cite{Langlands 1}), we explain why the scaling dimension-one operators in the holomorphic chiral algebra of the purely bosonic sector of the sigma-model will generate an affine $G$ OPE-algebra at the critical level. We then explain why the Casimir fields constructed out of these dimension-one currents must span the $\it{classical}$ holomorphic chiral algebra of the purely bosonic sector of the sigma-model, which, in turn, implies that their Laurent modes must generate the $\it{classical}$ centre $\frak{z}(\widehat{\frak g})$ of the completed universal enveloping algebra of the affine $G$-algebra at the critical level.

\newsubsection{The Twisted Sigma-Model  on $G/B$ and the Sheaf of CDO's}

As reviewed in appendix~A of \cite{Langlands 1}, the $\overline Q_+$-cohomology or the holomorphic chiral algebra of the twisted $(0,2)$ sigma-model on $X$ can be expressed in terms of the Cech-cohomology of the sheaf of CDO's. Since our main discussion involves the holomorphic chiral algebra of the sigma-model, and since we shall need to turn to some mathematical theorems regarding the sheaf of CDO's in our arguments, we shall first describe the sigma-model in terms of the sheaf of CDO's.

Recall that $X= G/B$, where $B$ is the subgroup of upper
triangular matrices of $G$ with a nilpotent Lie algebra $\frak
b$. Let us cover $X$ with $N$ open charts $U_w$ where
$w=1,2, \dots, N$, such that each open chart $U_w$ can be
identified with the affine space $\mathbb {C}^{n}$, where $n = \textrm{dim}_{\mathbb C}X$. Then, as explained in appendix~A of \cite{Langlands 1}, the sheaf of CDO's in any $U_w$---which describes a localised version of the sigma-model on $U_w$---can be described by $n$ free $\beta\gamma$ systems with the action \be I = \sum
_{i=1}^{n} \ {1\over 2\pi}\int|d^2z| \ \beta_i
\partial_{\bar z} \gamma^i. \label{beta-gamma action for G/B
on U1} \ee As before, the $\beta_i$'s and $\gamma^i$'s are fields
of dimension $(1,0)$ and $(0,0)$ respectively. They obey the
standard free-field OPE's; there are no singularities in the
operator products $\beta_i(z)\cdot \beta_i(z')$ and
$\gamma^i(z)\cdot\gamma^i(z')$, while \be \beta_i(z)\gamma^j(z')
\sim  -{\delta_i^j\over z-z'}. \ee

Similarly, the sheaf of CDO's in a neighbouring intersecting chart
$U_{w+1}$ is described by $n$ free
$\tilde\beta\tilde\gamma$ systems with action \be I= \sum
_{i=1}^{n}\ {1\over 2\pi}\int|d^2z| \ \tilde \beta_i
\partial_{\bar z} \tilde\gamma^i, \label{beta-gamma action for
G/B on U2} \ee where the $\tilde \beta_i$ and $\tilde
\gamma^i$ fields obey the same OPE's as the $\beta_i$ and
$\gamma^i$ fields. In other words, the non-trivial OPE's are given
by \be \tilde \beta_i(z) \tilde \gamma^j(z')  \sim
-{\delta_i^j\over z-z'}. \ee

In order to describe a globally-defined sheaf of CDO's, one will
need to glue the free conformal field theories with actions
(\ref{beta-gamma action for G/B on U1}) and (\ref{beta-gamma
action for G/B on U2}) in the overlap region $U_w \cap
U_{w+1}$ for every $w = 1,2, \dots N$, where $U_{1+ N} = U_1$.
To do so, one must use the admissible automorphisms of the free
conformal field theories defined in
(A.29)-(A.30) of \cite{Langlands 1} to glue the free-fields
together; they are given by
\begin{eqnarray}
\label{autoCDOgammaG}
{\tilde \gamma}^i & = & [g] ^i{}_j \ \gamma^j ,\\
\label{autoCDObetaG} {\tilde \beta}_i  & = &  \beta_k D^k{}_i
+
\partial_z \gamma^j E_{i j},
\end{eqnarray}
where $i,j,k = 1, 2, \dots, n$. Here, $g$, $D$ and $E$ are
$n \times n$ matrices, whereby $[g]$ is the matrix of geometrical transition functions,
$[(D^T)^{-1}]_i{}^k = \partial_i [g] ^k{}_j \gamma^j$ and $[E]_{ij} = \partial_i B_j$. It can be verified that
$\tilde \beta$ and $\tilde \gamma$ will obey the correct OPE's among
themselves~\cite{MSV}. Moreover, let $R_w$ represent a transformation of the
fields in going from $U_{w}$ to $U_{w+1}$. One can also verify
that  there is no anomaly to a global definition of a sheaf of
CDO's on $X =G/B$---a careful computation will reveal that
one will get the desired composition maps $(R_{N} \dots
R_4R_3R_2R_1) \cdot \gamma^j = \gamma^j$ and $ (R_{N}\dots
R_4R_3R_2R_1) \cdot \beta_i = \beta_i$. Again, this is just a
statement that one can always define a sheaf ${\widehat {\cal O}^{ch}_X}$ of CDO's  on any flag manifold $X = G/B$ \cite{MSV}. Physically, this just corresponds to the fact that since $p_1(X) =0$, the sigma-model will be well-defined and anomaly-free (see appendix~A of \cite{Langlands 1}).

\newsubsection{Global Sections of ${\widehat {\cal O}^{ch}_X}$ and an Affine $G$-algebra at the Critical Level}

Since $X=G/B$ is of complex dimension $n$, the chiral
algebra $\cal A$ of the sigma-model will be given by ${\cal A} = \bigoplus_{g_R
=0}^{g_R = {n} } H^{g_R}( X, {\widehat {\cal O}^{ch}_X})$
as a vector space. As in \cite{Langlands 1}, it would suffice for our purpose to
concentrate on the fermion-independent sector of $\cal A$---from
our $\overline Q_+$-Cech cohomology dictionary (explained in appendix~A of \cite{Langlands 1}), this again
translates to studying only the global sections in $H^0(X,
{\widehat {\cal O}^{ch}_X})$.

According to theorem 5.13 of \cite{MSV}, one can always find
elements in $H^0(X, {\widehat {\cal O}^{ch}_{X}})$ for any flag
manifold $X = G/B$, that will furnish a module of an affine
$G$-algebra at the critical level. This means that one can
always find dimension-one global sections of the sheaf ${\widehat
{\cal O}^{ch}_{X}}$ that correspond to $\psi^{\bar i}$-independent
currents $J^a(z)$ for $a =1,2, \dots \textrm{dim}({\frak
g})$, that satisfy the OPE's of an affine $G$-algebra at
the critical level $k = -h^{\vee}$: \be
 J_a (z) J_b (z') \sim -{{h^{\vee} d_{ab}}\over{(z-z')^2}} + \sum_c f_{ab}{}^c {{J_c(z')}\over {(z-z')}},
\ee where  $h^{\vee}$ is the dual Coxeter number of the Lie
algebra $\frak g$, and $d_{ab}$ is its Cartan-Killing
metric.\footnote{Note that one can consistently introduce
appropriate fluxes to deform the level away from $-h^{\vee}$---recall from our discussion in $\S$A.7 of \cite{Langlands 1} that the $E_{ij}= \partial_i B_j$ term in (\ref{autoCDObetaG}) is related to the fluxes
that correspond to the moduli of the chiral algebra, and since
this term will determine the level $k$ of the affine $G$-algebra via the identification of the global sections $\tilde
\beta_i$ with the affine currents valued in the subalgebra of
$\frak g$ associated to its positive roots, turning on the
relevant fluxes will deform $k$ away from $-h^{\vee}$. Henceforth,
whenever we consider $k\neq -h^{\vee}$, we really mean turning on
fluxes in this manner.} Since these current operators correspond to global sections, it will
be true that $\widetilde J_a(z) = J_a(z)$ on any $U_{w} \cap
U_{w+1}$ for all $a$, where $\widetilde J_a(z)$ and $J_a(z)$ are sections of the sheaf of CDO's defined in $U_w$ and $U_{w+1}$ respectively. Moreover, from our $\overline Q_+$-Cech
cohomology dictionary, they will be $\overline Q_+$-closed chiral
vertex operators that are holomorphic in $z$, which means that one
can expand them in a Laurent series that allows an affinisation of
the $G$ Lie-algebra generated by their resulting zero modes. The
space of these operators obeys all the physical axioms of a chiral
algebra except for reparameterisation invariance on the $z$-plane
or worldsheet $\Sigma$. We will substantiate this last statement
next by showing that the holomorphic stress tensor fails to exist
in the $\overline Q_+$-cohomology at the quantum level. Again,
this observation will be important in our discussion of a
geometric Langlands correspondence for $G$.

\newsubsection{The Segal-Sugawara Tensor and the Classical Holomorphic Chiral Algebra}

Recall that for any affine algebra $\widehat {\frak g}$ at
level $k \neq -h^{\vee}$, one can construct the corresponding
stress tensor out of the currents of $\widehat {\frak g}$ via
a Segal-Sugawara construction \cite{Ketov}: \be T(z) = {{:
d^{ab}J_aJ_b(z) :} \over {k+h^{\vee}}}. \label{SS def T(z) for
G}\ee As required, for every $k \neq {-h^{\vee}}$, the modes
of the Laurent expansion of $T(z)$ will span a Virasoro algebra.
In particular, $T(z)$ will generate holomorphic reparametrisations
of the coordinates on the worldsheet $\Sigma$. Notice that this
definition of $T(z)$ in (\ref{SS def T(z) for G}) is
ill-defined when $k=-h^{\vee}$. Nevertheless, one can always
associate $T(z)$ with the Segal-Sugawara operator $S(z)$ that is
well-defined at any finite level, whereby \be S(z) =  (k+h^{\vee})
T(z), \label{S(z) for G} \ee and \be S(z) = {: d^{ab} J_a J_b
(z):}. \label{s(z) for G} \ee From (\ref{S(z) for G}), we
see that $S(z)$ generates, in its OPE's with other field
operators, $(k+h^{\vee})$ times the transformations usually
generated by the stress tensor $T(z)$. Therefore, at the level $k=
-h^{\vee}$, $S(z)$ generates no transformations at all---its OPE's
with all other field operators are trivial. This is equivalent to
saying that the holomorphic stress tensor does not exist at the
quantum level, since $S(z)$, which is the only well-defined
operator at this level that could possibly generate the
transformation of fields under an arbitrary holomorphic
reparametrisation of the worldsheet coordinates on $\Sigma$, acts
by zero in the OPE's.

Despite the fact that $S(z)$ will cease to exist in the spectrum
of physical operators associated to the twisted sigma-model on
$X=G/B$ at the quantum level, it will nevertheless exist as a
field in its classical ${\overline Q}_+$-cohomology or holomorphic
chiral algebra. One can convince oneself that this is true as
follows. Firstly, from our ${\overline Q}_+$-Cech cohomology
dictionary, since the $J_a(z)$'s are in $H^0(X, {\widehat{\cal
O}}^{ch}_{X})$, it will mean that they are in the $\overline
Q_+$-cohomology of the sigma-model at the quantum level. Secondly,
since quantum corrections can only annihilate cohomology classes
and not create them, it will mean that the $J_a(z)$'s will be in
the classical $\overline Q_+$-cohomology of the sigma-model, i.e.,
the currents are $\overline Q_+$-closed and are therefore
invariant under the transformations generated by $\overline Q_+$
in the absence of quantum corrections. Hence, one can readily see
that $S(z)$ in (\ref{s(z) for G}) will also be $\overline
Q_+$-closed at the classical level. Lastly, recall from appendix~A of \cite{Langlands 1} that $[\overline Q _+, T(z)] = 0$ such that $T(z) \neq \{
{\overline Q}_+, \cdots \}$  in the absence of quantum corrections
to the action of $\overline Q_+$ in the classical theory. Note
also that the integer $h^{\vee}$ in the factor $(k+h^{\vee})$ of
the expression $S(z)$ in (\ref{S(z) for G}), is due to a shift
by $h^{\vee}$ in the level $k$ because of quantum renormalisation
effects \cite{Fuchs}, i.e., the classical expression of $S(z)$ for
a general level $k$ can actually be written as $S(z) = k T(z)$,
and therefore, one will have $[\overline Q_+, -h^{\vee} T(z)] =
[\overline Q_+, S(z)] = 0$, where $S(z) \neq \{ {\overline Q}_+,
\cdots \}$ in the classical theory. Therefore, $S(z)$ will be a
spin-two field in the classical holomorphic chiral algebra of the
purely bosonic sector of the twisted sigma-model on $X= G/B$.
This observation is also consistent with the fact that $S(z)$
fails to correspond to a global section of the sheaf ${\cal
O}^{ch}_{X}$ of CDO's---note that in our case, we actually have
$S(z) = -h^{\vee} T(z)$ in the classical theory, and this will
mean that under quantum corrections to the action of $\overline
Q_+$, we will have (see appendix~A of \cite{Langlands 1}) $[\overline Q_+, S(z)] = -h^{\vee}
\partial_z(R_{i \bar j}
\partial_z \phi^i \psi^{\bar j}) \neq 0$, since $R_{i\bar j} \neq 0$ for any flag manifold $G/B$. This corresponds in the Cech-cohomology picture to the expression ${\widetilde{\widehat S}(z)}
- {\widehat S}(z) \neq 0$ over an arbitrary intersection $U_w \cap U_{w+1}$ of open sets, where ${\widetilde{\widehat S}(z)}$ and ${{\widehat S}(z)}$ are sections of the sheaf of CDO's defined in $U_w$ and $U_{w+1}$ respectively. This means that ${\widehat S}(z)$, the Cech-cohomology counterpart to the $S(z)$ operator, will fail to be in
$H^0(X, {\widehat {\cal O}}^{ch}_{X})$. Consequently, one can
always represent $S(z)$ by a classical $c$-number. This point will
be important when we discuss how one can define Hecke eigensheaves
that will correspond to flat $^LG$-bundles on a Riemann surface
$\Sigma$ in our physical interpretation of the geometric Langlands
correspondence for $G$.

The fact that $S(z)$ acts trivially in any OPE with other field
operators implies that its Laurent modes will commute with the
Laurent modes of any of these other field operators; in
particular, they will commute with the Laurent modes of the
$J_a(z)$ currents---in other words, the Laurent modes of $S(z)$
will span the centre ${\frak z}(\widehat{\frak g})$ of the
completed universal enveloping algebra of the affine $G$-algebra $\widehat{\frak g}$ at the critical level
$k=-h^{\vee}$ (generated by the Laurent modes of the $J_a(z)$
currents in the quantum chiral algebra of the twisted sigma-model
on $G/B$ themselves).\footnote{Notice that $S(z)$ is constructed out of the
currents of the affine $G$-algebra by using the invariant
tensor $d^{ab}$ of the corresponding Lie algebra. Consequently, its Laurent
modes will span not the centre of the affine algebra, but rather
the centre of the completed universal enveloping algebra of the
affine algebra.} Notice also that $S(z)$ is $\psi^{\bar
j}$-independent and is therefore purely bosonic in nature. In
other words, the local field $S(z)$ exists only in the
$\it{classical}$ holomorphic chiral algebra of the $\psi^{\bar j}$-independent, $\it{purely}$
$\it{bosonic}$ sector of the twisted sigma-model on $X= G/B$.

\newsubsection{Higher-Spin Casimir Operators and the Classical Holomorphic Chiral Algebra}

For an affine $G$-algebra, one can generalise
the Sugawara formalism to construct higher-spin analogs of the
holomorphic stress tensor with the currents.  These higher-spin
analogs are called Casimir operators, and were first constructed
in \cite{casimir operators}.

In the context of an affine $G$-algebra with a module that is
furnished by the global sections of the sheaf of CDO's on
$X=G/B$, a spin-$s_i$ analog of the holomorphic stress tensor
will be given by the ${s_i}^{th}$-order Casimir operator
\cite{review}
 \be
T^{(s_i)} (z) = {{:{\tilde d}^{a_1 a_2 a_3 \dots a_{s_i}} (\frak g,k)
(J_{a_1} J_{a_2}\dots J_{a_{s_i}})(z):} \over {k+h^{\vee}}},
\label{casimir of spin-three for G)} \ee where ${\tilde
d}^{a_1a_2 a_3 \dots a_{s_i}}(\frak g, k)$ is a completely symmetric
traceless $\frak g$-invariant tensor of rank $s_i$ that
depends on the level $k$ of the affine $G$-algebra. It is also well-defined and finite at $k=-h^{\vee}$. The superscript on
$T^{(s_i)}(z)$ just denotes that it is a spin-$s_i$ analog of
$T(z)$. Note that $i = 1,2, \dots, \textrm{rank}(\frak g)$,
and  the spins $s_i$ can take the values $1+ e_i$, where $e_i$'s are the exponents of $\frak g$. Thus, one can have $\textrm{rank}(\frak g)$ of these Casimir operators, and the spin-2 Casimir operator is just
the holomorphic stress tensor $T(z)$ from the usual Sugawara construction.

As with $T(z)$ in (\ref{SS def T(z) for G}), $T^{(s_i)} (z)$
is ill-defined when $k = -h^{\vee}$. Nevertheless, one can always
make reference to a spin-$s_i$ analog of the Segal-Sugawara tensor
$S^{(s_i)}(z)$ that is well-defined for any finite value of $k$,
where its relation to $T^{(s_i)}(z)$ is given by \be S^{(s_i)}(z)
= (k+h^{\vee}) T^{(s_i)}(z), \ee and \be S^{(s_i)}(z) = :{\tilde
d}^{a_1 a_2 a_3 \dots a_{s_i}}(\frak g, k) (J_{a_1} J_{a_2}\dots
J_{a_{s_i}})  (z):. \label{S^{(s_i)}(z)} \ee That is, the operator
$S^{(s_i)}(z)$ generates in its OPE's with all other operators of
the quantum theory,  $(k+h^{\vee})$ times the field
transformations generated by $T^{(s_i)}(z)$.

Notice  however, that at $k=-h^{\vee}$, $S^{(s_i)}(z)$ acts by
zero in its OPE with any other operator. This is equivalent to
saying that $T^{(s_i)}(z)$ does not exist as a quantum operator, since the only well-defined operator $S^{(s_i)}(z)$ which is supposed to generate the field transformations associated to
$T^{(s_i)}(z)$, act by zero and thus generate no field
transformations at all. From our $\overline Q_+$-Cech cohomology
dictionary, this means that the $\psi^{\bar i}$-independent
operator $T^{(s_i)}(z)$ will fail to correspond to a dimension
$s_i$ global section of $\widehat{\cal O}^{ch}_X$. Since we have,
at the classical level, the relation $S^{(s_i)}(z) = -h^{\vee}
T^{(s_i)}(z)$, it will mean that $S^{(s_i)}(z)$ will also fail to
correspond to a dimension $s_i$ global section of $\widehat{\cal
O}^{ch}_X$. Thus, $S^{(s_i)}(z) $ will fail to be an operator at
the quantum level. Is it even a spin-$s_i$ field in the classical
holomorphic chiral algebra of the twisted sigma-model on
$G/B$, one might ask. The answer is ``yes''. To see this, recall
that each of the $J_{a_k}(z)$'s are separately $\overline
Q_+$-invariant and not $\overline Q_+$-exact at the classical
level. Therefore, the classical counterpart of $S^{(s_i)}(z)$ in
(\ref{S^{(s_i)}(z)}) must also be such, which in turn means that
it will be in the classical $\overline Q_+$-cohomology and hence
classical holomorphic chiral algebra of the twisted sigma-model on $G/B$.

The fact that the $S^{(s_i)}(z)$'s act trivially in any OPE with
other field operators implies that their Laurent modes will
commute with the Laurent modes of any other operator; in
particular, they will commute with the Laurent modes of the
currents $J_a(z)$ for $a=1,2, \dots, \textrm{dim}(\frak g)$---in other words, the Laurent modes of all $\textrm{rank}(\frak g)$ of the $S^{(s_i)}(z)$ fields will span fully the
centre ${\frak z}(\widehat{\frak g})$ of the completed
universal enveloping algebra of $\widehat{\frak g}$ at the
critical level $k=-h^{\vee}$ (generated by the
Laurent modes of the $J_a(z)$ currents of the quantum chiral
algebra of the twisted sigma-model on $G/B$ themselves). Last but not
least, notice that the $S^{(s_i)}(z)$ fields are also $\psi^{\bar
j}$-independent and are therefore purely bosonic in nature. In
other words, the local fields $S^{(s_i)}(z)$, for $i=1,2,\dots
\textrm{rank}(\frak g)$---whose Laurent modes will together
generate ${\frak z}(\widehat{\frak g})$---exist only in the
$\it{classical}$ holomorphic chiral algebra of the $\psi^{\bar j}$-independent, $\it{purely}$
$\it{bosonic}$ sector of the
twisted sigma-model on $X=G/B$.

\newsubsection{The Centre ${\frak z}(\widehat{\frak g})$ as a Poisson Algebra ${\cal W}_{-h^{\vee}}(\widehat{\frak g})$}

For an affine $G$-algebra at an arbitrary level $k \neq
-h^{\vee}$, the $S^{(s_i)}(z)$'s will exist as $\psi^{\bar j}$-independent quantum operators in the $\overline Q_+$-cohomology of the sigma-model.  According to our $\overline Q_+$-Cech cohomology dictionary, the $S^{(s_i)}(z)$'s then correspond to classes in $H^0(X, \widehat{\cal O}^{ch}_X)$, Since the cup product of sheaf cohomologies map products of global sections to global sections, it will mean that the OPE of any two $S^{(s_i)}(z)$ operators must contain another $S^{(s_i)}(z)$ operator. Moreover, since all the $S^{(s_i)}(z)$ operators are $\overline Q_+$-closed, they must form a closed OPE-algebra.\footnote{Note that if $\cal O$ and $\cal O'$ are non-exact $\overline
Q_+$-closed observables in the $\overline
Q_+$-cohomology, i.e., ${\{\overline Q_{+}, \cal O\}}={\{\overline
Q_{+}, {\cal O}'\}}= 0$, then $\{\overline Q_{+}, {\cal O}{\cal
O}'\} =0$. Moreover, if $\{\overline Q_{+}, {\cal O}\}=0$, then
${\cal O}\{\overline Q_{+}, W\}= \{\overline Q_{+}, {\cal O}W\}$
for any observable $W$.  These two statements mean that the
cohomology classes of observables that commute with $\overline
Q_{+}$ form a closed and well-defined algebra.} What then is this closed OPE-algebra?

To answer this, first recall that for some $k \neq -h^{\vee}$, the $S^{(s_i)}(z)$'s have a quantum definition whereby $S^{(s_i)}(z) = (k+h^{\vee}) T^{(s_i)}(z)$. The Casimir operators $T^{(s_i)}(z)$ are know to span (up to null or $\overline Q_+$-exact operators in our interpretation) a closed $\cal W$ OPE-algebra associated to $\widehat {\frak g}$ \cite{review}. Since the spin-2 Casimir operator $T^{(2)}(z)$ generates a Virasoro OPE-algebra of central charge $c = k \ \textrm{dim}(\frak g) /(k + h^{\vee})$, the $S^{(s_i)}(z)$'s will then span a rescaled (by a factor of $(k + h^{\vee})$) version of the closed $\cal W$ OPE-algebra associated to $\widehat {\frak g}$ of central charge $c = k \
\textrm{dim}(\frak g)  / (k+h^{\vee})$ for $k\neq -h^{\vee}$.

Since each $S^{(s_i)}(z)$ is holomorphic in $z$, we can Laurent
expand it as \be S^{(s_i)}(z) = \sum_{n \in {\mathbb Z}} {\hat
S}^{(s_i)}_n z^{-n-{s_i}}. \label{S^{(s_i)}(z) quantum for G}
\ee Let us henceforth denote ${\cal W}_k (\widehat{\frak g})$ to be the closed algebra generated by the Laurent modes ${\hat S}^{(s_i)}_n$ where $k \neq -h^{\vee}$. At $k \neq -h^{\vee}$, since $S^{(2)}(z) = (k+h^{\vee}) T(z)$, the Laurent modes ${\hat S}^{(2)}_n$ must then generate the
Virasoro algebra with the following quantum commutator relations: \be {[{\hat S}^{(2)}_n, {\hat
S}^{(2)}_m ]}  =  (k+ h^{\vee}) \left( (n-m) {\hat S}^{(2)}_{n+m}
+ { k \ \textrm{dim}({\frak g})  \over 12} \  (n^3-n) \
\delta_{n, -m}\right). \label{S^{(2)}(z) quantum for G} \ee
Likewise, the other quantum commutator relations spanned by the Laurent modes of the other spin-$s_i$ operators, will take the same form as (\ref{S^{(2)}(z) quantum for G}) and have a factor of $(k+h^{\vee})$ in front. Since we will have no need to refer to these explicit relations in our discussions, we shall omit them for brevity, as they can get rather complicated very quickly.

Now, let us consider the case when $k = -h^{\vee}$. From our
earlier explanations about the nature of the $S^{(s_i)}(z)$
operators, we find that they will cease to exist as quantum
operators at this critical level. Since we understand that the
$S^{(s_i)}(z)$'s must be  holomorphic classical fields at $k
=-h^{\vee}$, we shall rewrite the Laurent expansion of
$S^{(s_i)}(z)$ as \be S^{(s_i)}(z) = \sum_{n \in {\mathbb Z}}
S^{(s_i)}_n z^{-n-{s_i}}, \label{S^{(s_i)}(z) classical for SL(N)}
\ee so as to differentiate the classical modes of expansion
$S^{(s_i)}_n$ from their quantum counterparts ${\hat S}^{(s_i)}_n$
in (\ref{S^{(s_i)}(z) quantum for G}). Unlike the ${\hat
S}^{(s_i)}_n$'s which obey the quantum commutator relations of a
${\cal W}_k (\widehat{\frak g})$-algebra for an arbitrary level $k \neq -h^{\vee}$,
the $S^{(s_i)}_n$'s, being the modes of a Laurent expansion of a
classical field, will instead obey Poisson bracket relations that
define a certain classical algebra at $k=-h^{\vee}$. Since every
${\hat S}^{(s_i)}_n$ must reduce to its classical counterpart
${S}^{(s_i)}_n$ at $k = -h^{\vee}$,  one can see that by taking
$(k+h^{\vee}) \to 0$, we are actually going to the classical
limit. This is analogous to taking the  ${\hbar} \to 0$ limit in
any quantum mechanical theory whenever one wants to ascertain its
classical counterpart. In fact, by identifying $(k+h^{\vee})$ with
$i \hbar$, and by noting that one must make the replacement from
Possion brackets to commutators---that is, $\{E^{(s_i)}_n, E^{(s_j)}_m
\}_{P.B.} \rightarrow {1\over {i \hbar}} [ {\hat E}^{(s_i)}_n,
{\hat E}^{(s_j)}_m ]$---in quantising any classical mode
$E^{(s_i)}_n$ into an operator  ${\hat E}^{(s_i)}_n$, we can
ascertain the classical algebra generated by the $S^{(s_i)}_n$'s
from the ${\cal W}_{k} (\widehat{\frak g})$-algebra commutator relations that their quantum counterparts---the ${\hat S}^{(s_i)}_n$'s---satisfy. Since all the $S^{(s_i)}(z)$ fields must now lie in the classical
$\overline Q_+$-cohomology of the twisted sigma-model on
$G/B$, it will mean that their Laurent modes $S^{(s_i)}_n$
 must also generate a closed, classical algebra associated to $\widehat{\frak g}$, which, we shall henceforth denote as ${\cal W}_{-h^{\vee}} (\widehat{\frak g})$. In order to
ascertain the central charge of this $\it{classical}$ ${\cal W}_{-h^{\vee}} (\widehat{\frak g})$-algebra, it suffices to determine the central charge of its classical Virasoro subalgebra generated by the $S^{(2)}_m$'s. From
(\ref{S^{(2)}(z) quantum for G}), we find  that as $k \to
-h^{\vee}$, the $S^{(2)}_m$'s satisfy \be \{{S}^{(2)}_n,
{S}^{(2)}_m \}_{P.B.}  =   (n-m) {S}^{(2)}_{n+m} - { h^{\vee} \
\textrm{dim}({\frak g})  \over 12} \  (n^3-n) \ \delta_{n,
-m}, \label{S^{(2)}(z) classical for G} \ee the classical
Virasoro algebra with central charge $c = - h^{\vee}
\textrm{dim}(\frak g)$. Hence, the $S^{(s_i)}_n$'s will
generate a classical ${\cal W}_{-h^{\vee}} (\widehat{\frak g})$-algebra of central charge $c =
- h^{\vee} \textrm{dim}(\frak g)$. For example, the specific case of
$\frak g= \frak{sl}_2$ was considered in $\S$2.1 of \cite{Langlands 1}---the modes
$S^{(2)}_m$ were found to generate a classical ${\cal W}_{-h^{\vee}} (\widehat{\frak {sl}}_2)$-algebra
with central charge $c= - h^{\vee} \textrm{dim}(\frak {sl}_2)
=-6$, where $h^{\vee} =2$ and $ \textrm{dim}(\frak {sl}_2) =3$.
The specific case of $\frak g= \frak{sl}_3$ was also considered in
$\S$2.3 of \cite{Langlands 1}---the modes $S^{(2)}_m$ and $S^{(3)}_m$ were found to
generate a classical ${\cal W}_{-h^{\vee}} (\widehat{\frak {sl}}_3)$-algebra with central charge $c=
- h^{\vee} \textrm{dim}(\frak {sl}_3) =-24$, where $h^{\vee} = 3$
and $ \textrm{dim}(\frak {sl}_3) =8$.

Recall at this point that the Laurent modes of the
$S^{(s_i)}(z)$ fields for $i=1, 2, \dots \textrm{rank}(\frak g)$, will together generate
$\frak{z}(\widehat{\frak g})$, the centre of the completed
universal enveloping algebra of the affine $G$-algebra
$\widehat{\frak g}$ at the critical level $k = -h^{\vee}$. Hence, we have an identification of Poisson algebras $\frak{z}(\widehat{\frak g}) \simeq {\cal W}_{-h^{\vee}} (\widehat{\frak g})$.

Last but not least, another way to understand why $\frak{z}(\widehat{\frak g})$ must be a classical (or Poisson) algebra is as follows. Firstly, let us consider the general case of $k \neq -h^{\vee}$, whereby the ${\hat S}^{(2)}_n$ modes can be related to the $J^a_n$ modes of $\widehat{\frak g}$ via the quantum commutator relations
\begin{eqnarray}
[{\hat S}^{(2)}_n, J^a_m] & = & -(k + h^{\vee}) m J^a_{n+m},\\
{[{\hat S}^{(2)}_n, {\hat S}^{(2)}_m ]} & = & (k+ h^{\vee}) \left( (n-m) {\hat
S}^{(2)}_{n+m} + {k \over 12} \ \textrm{dim}({\frak g})\ (n^3-n) \
\delta_{n, -m}\right), \label{S_n}
\end{eqnarray}
where $a = 1, 2, \dots, \textrm{dim}({\frak g})$. If we now let $k=-
h^{\vee}$, we will have $[{\hat S}^{(2)}_n, J^a_m] = [{\hat S}^{(2)}_n, {\hat S}^{(2)}_m] =0$. Hence, one can define
simultaneous eigenstates of the ${\hat S}^{(2)}_n$ and $J^a_n$ mode
operators. In particular, one would be able to properly define a
general state ${\Psi} = {\hat S}^{(2)}_{-l} {\hat S}^{(2)}_{-q} \dots {\hat
S}^{(2)}_{-p} |0, \alpha \rangle$, where $| 0, \alpha \rangle$ is a
vacuum state associated to a representation of $\frak g$
labelled by $\alpha$, such that $J^a_0 | 0, \alpha \rangle =
\alpha^a | 0, \alpha \rangle$. However, note that any such $\Psi$ will
correspond to a null-state, i.e., $\Psi$ decouples from the real,
physical Hilbert space of quantum states spanned by the
representations of $\frak g$ \cite{lindstrom}. This means
that the ${\hat S}^{(2)}_{m} $'s which span ${\frak
z}(\widehat{\frak g})$ cannot exist as quantum mode operators.
Hence, since ${\frak z}(\widehat{\frak g})$ must be closed in all the ${\hat S}^{(s_i)}_{m} $ modes, it must therefore be a classical algebra at $k = - h^{\vee}$.

\newsection{The $B$-gauged WZW model on $G$ and the $\cal W$-Algebra ${\cal W}_{k'}(\widehat {\frak g})$}

In this section, we shall explain how a $\it{dual}$ description of the holomorphic chiral algebra of the purely bosonic sector of the sigma-model on $G/B$, can be given by the $\it{holomorphic}$ BRST-cohomology (or chiral
algebra) of a $B$-gauged WZW model on $G$. We then proceed to outline the mathematical Drinfeld-Sokolov reduction procedure \cite{book} of generating the Hecke algebra of local operators which defines ${\cal
W}_{k'}(\widehat {\frak g})$---the $\cal W$-algebra associated to
${\widehat {\frak g}}$ at level $k'$. Lastly, we will show that the holomorphic BRST-cohomology of the $B$-gauged WZW model on $G$  at level $k'$ physically realises, in all generality, this Hecke algebra of local operators.

\newsubsection{A Dual Description of the Purely Bosonic Holomorphic Chiral Algebra}

Let us now seek a dual description of the above classical,
holomorphic chiral algebra of the twisted sigma-model on $G/B$
spanned by the $S^{(s_i)}(z)$'s. To this end, let us first generalise the action
of the twisted sigma-model by making the replacement $g_{i \bar j}
\to g_{i \bar j} + b_{i \bar j}$ in $V$ of $S_{\textrm twist}$ of
(A.9) in \cite{Langlands 1}, where $b_{i \bar j}$ is a $(1,1)$-form on the
target space $X$ associated to a B-field. This just adds to
$S_{\textrm{twist}}$ a $\it{cohomologically}$-$\it{trivial}$
$\overline Q_+$-exact term $\{\overline Q_+, - b_{i \bar j}
\psi^i_{\bar z} \partial_z \phi^{\bar j} \}$, and does nothing to change our above
discussions about the chiral algebra of the sigma-model.
This generalised action can be explicitly written as
\begin{eqnarray}
S_{\mathrm gen}& = & \int_{\Sigma} |d^2z| \  (g_{i{\bar j}} + b_{i
\bar j})(
\partial_z \phi^{\bar j} \partial_{\bar z}\phi^i) + g_{i \bar j}
\psi_{\bar z}^i {D}_z \psi^{\bar j} + b_{i \bar j} \psi_{\bar z}^i
{\partial}_z \psi^{\bar j} + b_{{i \bar l},{\bar j}} \psi^i_{\bar
z}\partial _z \phi^{\bar l} \psi^{\bar j}. \label{actiongen}
\end{eqnarray}

Now recall that the $S^{(s_i)}(z)$'s exist in the classical holomorphic chiral
algebra of the $\psi^{\bar j}$-independent sector of the twisted sigma-model on $G/B$. This means that in order
for one to ascertain the dual description of the $S^{(s_i)}(z)$'s, it suffices
to confine oneself to the study of the holomorphic chiral algebra
of the $\psi^{\bar j}$-independent, purely bosonic sector of the twisted
sigma-model on $G/B$. The $\psi^{\bar
j}$-independent specialisation of $S_{\textrm{gen}}$, which
describes this particular sector of interest, can be written as
\be S_{\textrm{bosonic}} = \int_{\Sigma} |d^2z| \ (g_{i\bar j} +
b_{i \bar j}) \partial_{\bar z} \phi^i \partial_z \phi^{\bar
j}.\label{Sbosonic explicit} \ee

Notice that $S_{\textrm{bosonic}}$ just describes a non-linear sigma-model of a free bosonic string which
propagates in a $G/B$ target-space. Note that a non-linear sigma-model on any
homogenous coset space such as $G/B$, can be described by an
$\it{asymmetrically}$ $B$-gauged WZW model on $G$ that is associated with the
action $g \to gb^{-1}$, where $g \in G$ and $b \in B$. However, upon a BRST-quantisation, one can easily see that the BRST-cohomology of the asymmetrically $B$-gauged WZW model on $G$, coincides exactly with the $\it{holomorphic}$ (or left-moving) sector of the total BRST-cohomology of a
$\it{symmetrically}$ $B$-gauged WZW model on $G$ that is associated with the
action $g \to bgb^{-1}$. In other words, at the level of the holomorphic chiral algebra, a physically equivalent description of the $\psi^{\bar j}$-independent, non-supersymmetric sector of the twisted sigma-model on $G/B$, will be given by a symmetrically $B$-gauged WZW model on $G$ that is genuinely gauge-invariant on the worldsheet $\Sigma$.\footnote{Henceforth, whenever we refer to the $B$-gauged WZW model on $G$, we really mean the symmetrically gauged $WZW$ model on $G$ that is genuinely gauge-invariant on the worldsheet $\Sigma$.} In other words, the $S^{(s_i)}(z)$'s should correspond to observables in the classical holomorphic BRST-cohomology of the $B$-gauged WZW model on
$G$.

\newsubsection{The $B$-gauged WZW Model on $G$}

let us now proceed to describe the relevant $B$-gauged WZW model on $G$ in detail.\footnote{It may be disconcerting to
some readers at this point that the Borel subgroup $B$ which we
are gauging the $G$ WZW model by, is non-compact in
general. Apart from citing several well-known examples in the
physics literature \cite{review 23, review 24, sourdis,O&B, ref
for gauged WZW} that have done likewise to consider non-compact
WZW models gauged to non-compact (sometimes Borel) subgroups, one
can also argue that our model is actually equivalent---within our
context---to a physically consistent model which gauges a
$\it{compact}$ subgroup instead. Firstly, note that for a complex
flag manifold $G/B$, we have the relation $G/B = {\cal G}/ {\cal T}$, where ${\cal G}$ is a compact group whose Lie algebra is the real form of the Lie algebra of $G$, and $\cal T$ is the maximal torus of purely diagonal matrices in
$\cal G$~\cite{intro to lie groups}---in other words, $\cal T$ is an
anomaly-free, $\it{compact}$ diagonal subgroup in the context of a
$\cal T$-gauged WZW model on $\cal G$. Secondly, note that the OPE
algebras of the affine $G$-algebra and the affine $\cal G$-algebra are the same. These two points imply that at the level of their holomorphic BRST-cohomologies, the $B$-gauged WZW model on $G$ is equivalent to the $\cal T$-gauged WZW model on $\cal G$ that can always be physically consistently defined, and whose gauge group is
also compact. However, since one of our main aims in this paper is to
relate the gauged WZW model to the algebraic DS-reduction scheme, we want to consider the B-gauged WZW model on
$G$. Last but not least, note that we will ultimately
be interested only in the $\it{classical}$ spectrum of the gauged WZW
model, whereby the compactness or non-compactness of the
gauge group will be irrelevant.} First, note that the action of the most general WZW model can be written as
\be S_{\textrm{WZ}}(g) = { k ' \over {4 \pi}} \int_{\Sigma} d^2z \  \textrm{Tr}
(\partial_{z} g^{-1} \partial_{\bar z} g)  + { i k ' \over {24
\pi}} \int_{B; \partial B = \Sigma} d^3 x \ \textrm{Tr} (g^{-1} d
g)^3, \label{WZW action}
 \ee
where $k'$ is the level, and $g$ is a worldsheet scalar field
valued in any connected Lie group $G$ that is also periodic along one of the worldsheet
directions with period $2 \pi$.\footnote{Note that in some situations, the target group manifold of the WZW model is not simply-connected; the complex $D$-group or $SO(N, \bf{C})$ manifold is one such example. In this case, the non-simple-connectedness of the group will translate to a restriction in the values that $k'$ can take \cite{Polchinski 2}. In other situations, one must exclude some representations and include winding sectors in the Hilbert space of states. However, since our results will only depend on the classical spectrum of local fields of the WZW model in the limit $k' \to \infty$, we can, for our purpose, ignore this technical subtlety.} The trace $\textrm{Tr}$ is the
usual matrix trace in the defining representation of $G$.

A gauged version of (\ref{WZW action}) can be written as
\begin{eqnarray}
S_{\textrm{gauged}} (g, A_z, A_{\bar z}) & = & S_{\textrm{WZ}} (g)  + {k' \over {2\pi}}\int_{\Sigma} d^2z \  \textrm{Tr}  [ A_z (\partial_{\bar z} g g^{-1} + {\bar M}) -  A_{\bar z}(g^{-1} \partial_z g + {M})   \nonumber \\
&& \hspace{3.0cm}  + A_z g A_{\bar z} g^{-1} - A_z A_{\bar z}],
\label{gauged WZW action}
\end{eqnarray}
where the worldsheet one-form gauge field $A= A_z dz + A_{\bar z}
d{\bar z}$ is valued in $\frak h$, the Lie algebra of a subgroup
$H$ of $G$. Notice that $S_{\textrm{gauged}} (g, A_z, A_{\bar z})$
differs slightly from the standard form of a gauged WZW model
commonly found in the physical literature---additional $\bar M$
and $M$ constant matrices have been incorporated in the
$\partial_{\bar z} g g^{-1}$ and $g^{-1} \partial_z g $ terms of
the standard action, so that one can later use them to derive the
correct form of the holomorphic stress tensor without reference to
a coset formalism. Setting $\bar M$ and $M$ to the zero matrices
simply takes us back to the standard action for the gauged WZW
model. As required, $S_{\textrm{gauged}} (g, A_z, A_{\bar z})$ is
invariant under the standard (chiral) local gauge transformations
\be g \to hgh^{-1}; \ \ \ A_z \to
\partial_z h \cdot h^{-1} + h A_z h^{-1}; \ \ \ A_{\bar z} \to
\partial_{\bar z} h \cdot h^{-1} + h A_{\bar z} h^{-1},
\label{gauge tx} \ee where $h = e^{\lambda (z, \bar z)} \in H$ for
any $\lambda (z, \bar z) \in {\frak h}$.\footnote{A similar model
has been considered in \cite{ref for gauged WZW}. However, the
action in that context is instead invariant under a
$\it{non}$-$\it{chiral}$ local gauge transformation. Moreover, it
does not contain the $A_zA_{\bar z}$ term present in a standard
gauged WZW model.} The invariance of (\ref{gauged WZW action})
under the gauge transformations in (\ref{gauge tx}) can be
verified as follows. Firstly, note that the $\bar
M(M)$-independent terms make up the usual Lagrangian for the
standard gauged WZW action, which is certainly invariant under the
gauge transformations of (\ref{gauge tx}). Next, note that under
an infinitesimal gauge transformation $h \simeq 1+ \lambda$, the
terms $\textrm{Tr} (A_{z} \ \bar M)$ and $\textrm{Tr} (A_{\bar z}
\ M)$ change as
\begin{eqnarray} \label{variation Tr 1}\delta \textrm{Tr} (A_{z}
\ \bar M) & = & \textrm{Tr} (\partial_{z} \lambda \
\bar M) - \textrm{Tr} (\bar M \ [\lambda, A_z]),  \\
\delta \textrm{Tr} (A_{\bar z} \ M) & = & \textrm{Tr}
(\partial_{\bar z} \lambda \ M) - \textrm{Tr} (M \ [\lambda,
A_{\bar z}]). \label{variation Tr} \end{eqnarray} Since we will be
considering the case where $H$ is the Borel subgroup of $G$ and
therefore, $\lambda$ and $A$ will be valued in the Lie algebra of
a maximally $\it{solvable}$ (Borel) subgroup of $G$, the second
term on the R.H.S. of (\ref{variation Tr 1}) and (\ref{variation
Tr}) will be zero \cite{ref for gauged WZW}. What  remains are
total divergence terms that will vanish upon integration on
$\Sigma$ because it is a worldsheet with no boundaries. Therefore,
unless $H$ is a Borel subgroup of $G$ (or any other solvable
subgroup of $G$), one cannot incorporate $\bar M$ and $M$ in the
action and still maintain the requisite gauge invariance. This
explains why generalisations of gauged WZW models with these
constant matrices $\bar M$ and $M$ have not appeared much in the
physical literature. Nevertheless, this generalisation can be
considered in our case. As we shall see shortly, this
generalisation will allow us to obtain the correct form of the
holomorphic stress tensor of the $B$-gauged WZW model on $G$
without any explicit reference to a coset formalism.

The classical equations of motion that follow from the field
variations in (\ref{gauge tx}) are
\begin{eqnarray}
\label{1}
\delta A_z & : & D_{\bar z} g g^{-1}|_H = -M_+, \\
\label{2}
\delta A_{\bar z} & : & g^{-1} D_z g |_H = -M-, \\
\label{3}
\delta g & : & D_{\bar z} (g^{-1} D_z g ) = F_{z \bar z}, \\
\label{4} \delta g & : & D_{z} (D_{\bar z} g g^{-1}) = F_{\bar z
z},
\end{eqnarray}
where $F_{z \bar z} = \partial_z A_{\bar z} - \partial_{\bar z}
A_z + [A_z, A_{\bar z}]$ and $F_{\bar z z} = \partial_{\bar z} A_z
- \partial_{z} A_{\bar z} + [A_{\bar z}, A_{z}]$ are the
non-vanishing components of the field strength, and the covariant
derivatives are given by $D_z =
\partial_z + [A_z, \ ]$ and $D_{\bar z} = \partial_{\bar z} +
[A_{\bar z}, \ ]$. By imposing the condition of (\ref{2}) in
(\ref{3}), and by imposing the condition of (\ref{1}) in
(\ref{4}), since $M_{\pm}$ are constant matrices, we find that we
have the zero curvature condition $F_{z \bar z} = F_{\bar z z} =
0$ as expected of a non-dynamically gauged WZW model. This means
that $A_z$ and $A_{\bar z}$ are trivial on-shell. One is then free
to use the gauge invariance to set $A_z$ and/or $A_{\bar z}$ to a
constant such as zero. In setting $A_z = A_{\bar z} =0$ in
(\ref{3}) and (\ref{4}), noting that $F_{z \bar z}= F_{\bar z z} =
0$, we have the relations \be
\partial_{\bar z} (g^{-1} \partial_z g ) = 0 \qquad \textrm{and} \qquad \partial_{z} (\partial_{\bar z} g g^{-1}) =
0. \label{con} \ee In other words, we have a $\frak g$-valued,
holomorphic conserved current $J(z) = g^{-1} \partial_z g$, and a
$\frak g$-valued antiholomorphic conserved current $\bar J(\bar z)
= \partial_{\bar z} g g^{-1}$, both of which are dimension one and
generate affine symmetries on $\Sigma$. The action in (\ref{gauged
WZW action}) can thus be written as \begin{eqnarray} \label{gauged
WZW action simplified}
S_{\textrm{gauged}} (g, A_z, A_{\bar z}) & = & S_{\textrm{WZ}} (g)  + {k' \over {2\pi}}\int_{\Sigma} d^2z \  \textrm{Tr}  [ A_z ({\bar J}(\bar z) + \bar M) -  A_{\bar z}(J(z) + M)   \nonumber \\
&& \hspace{3.0cm}  + A_z g A_{\bar z} g^{-1} - A_z A_{\bar z}].
\end{eqnarray}

For our case where $H$ is a Borel subgroup $B$ of $G$, one can
further simplify (\ref{gauged WZW action simplified}) as follows.
Firstly, since $G$ is a connected group, its Lie algebra $\frak g$ will have a
Cartan decomposition ${\frak g} = {\frak n}_- \oplus {\frak c}
\oplus {\frak n}_+$, where $\frak c$ is the Cartan subalgebra, and
${\frak n}_{\pm}$ are the nilpotent subalgebras of the the upper
and lower triangular matrices of $G$. The Borel subalgebras will
then be given by ${\frak b}_{\pm} = {\frak c} \oplus {\frak
n}_{\pm}$, and they correspond to the Borel subgroups $B_{\pm}$.
For the complex flag manifolds that we will be considering in this
paper, $B_+$ will be the Borel subgroup of interest. $B$ will
henceforth mean $B_+$ in all of our proceeding discussions. With
respect to this decomposition of the Lie algebra $\frak g$, we can
write $ J(z) = \sum_{a=1}^{\textrm{dim}{{\frak n}_-}} J^a_- (z)
t^{-}_a + \sum_{a=1}^{\textrm{dim}{{\frak c}}} J^a_c(z) t^{c}_a +
\sum_{a=1}^{\textrm{dim}{{\frak n}_+}} J^a_+(z) t^{+}_a$, and $
{\bar J}(\bar z) = \sum_{a=1}^{\textrm{dim}{{\frak n}_-}} {\bar
J}^a_- (\bar z) t^{-}_a + \sum_{a=1}^{\textrm{dim}{{\frak c}}}
{\bar J}^a_c(\bar z) t^{c}_a + \sum_{a=1}^{\textrm{dim}{{\frak
n}_+}} {\bar J}^a_+(\bar z) t^{+}_a$, where $t^{-}_a \in {\frak
n}_-$, $t^{c}_a \in {\frak c}$, and $t^{+}_a \in {\frak n}_+$. One
can also write $M = \sum_{a=1}^{\textrm{dim}{{\frak n}_-}} M^a_-
t^{-}_a + \sum_{a=1}^{\textrm{dim}{{\frak c}}} M^a_c t^{c}_a  +
\sum_{a=1}^{\textrm{dim}{{\frak n}_+}} M^a_+ t^{+}_a$, and $\bar M
= \sum_{a=1}^{\textrm{dim}{{\frak n}_-}} {\bar M}^a_- t^{-}_a +
\sum_{a=1}^{\textrm{dim}{{\frak c}}} {\bar M}^a_c t^{c}_a  +
\sum_{a=1}^{\textrm{dim}{{\frak n}_+}} {\bar M}^a_+ t^{+}_a$,
where $M^a_{\pm ; c}$(${\bar M}^a_{\pm ; c}$) are arbitrary number
constants. Next, note that $H=B$, and $B \simeq N_+$, where $N_+ =
[B,B]$ is the subgroup of $G$ generated by its Lie algebra ${\frak
n}_+$ of strictly upper triangular matrices which are traceless,
i.e., for $t,t' \in {\frak n}_+$, we have $\textrm{Tr}_{L} (tt') -
\textrm{Tr}_{R}(t't) = 0$, where the trace $\textrm{Tr}_{L}$ and
$\textrm{Tr}_{R}$ are taken over some $L$ and $R$ representation
of $G$ respectively. In other words, $N_+$ is the non-anomalous
subgroup to be gauged, and we can write $A_{z} = \sum_{a
=1}^{\textrm{dim}{\frak n}_+} {\tilde A}_{z}^a t^+_a$, and
$A_{\bar z} =   \sum_{a =1}^{\textrm{dim}{\frak n}_+} {\tilde
A}_{\bar z}^a t^+_a$. Next, note that since $\textrm{Tr}
(t^{\alpha}_a t^{\beta}_b) = \delta_{a,b}\delta^{\alpha, \beta}$,
the trace of the second term on the R.H.S. of (\ref{gauged WZW
action simplified}) will be non-vanishing only  for components of
$J(z)$($\bar J(\bar z)$) and $M$($\bar M$) that are associated to
their expansion in ${\frak n}_+$. Let us denote ${J}^+(z) =
\sum_{a=1}^{\textrm{dim}{\frak n}_+} {J}^a_+ (z) t^{+}_a$ and
${M}^+ = \sum_{a=1}^{\textrm{dim} {{\frak n}_+}} {M}^a_+ t^{+}_a$.
Let us also denote ${\bar J}^+(\bar z) =
\sum_{a=1}^{\textrm{dim}{\frak n}_+} {\bar J}^a_+ (\bar z)
t^{+}_a$ and ${\bar M}^+ = \sum_{a=1}^{\textrm{dim} {{\frak n}_+}}
{\bar M}^a_+ t^{+}_a$. Then, one can write the action of a
$B$-gauged WZW model on $G$ as
\begin{eqnarray} S_{\textrm{B-gauged}} (g, A_{z}, A_{\bar z}, J^+,
{\bar J}^+)& = & S_{\textrm{WZ}} (g) - {k' \over {2\pi}} \int_{\Sigma} d^2z \
\sum_{l=1}^{\textrm{dim}{{\frak n}_+}} \left[ {\tilde A}^l_{\bar z}( J^l_+(z) + {M}^l_+) -
{\tilde A}^l_{z}( \bar J^l_+(\bar z) + {\bar
M}^l_+) \right]\nonumber \\
&& \hspace{4.5cm} - \textrm {Tr} [ {A}_z g {A}_{\bar z} g^{-1} -
{A}_z {A}_{\bar z}].
\label{B-gauged WZW action G}
\end{eqnarray}

Due to the $B$-gauge invariance of the theory, we must divide the
measure in any path integral computation by the volume of the
$B$-gauge symmetry. That is, the partition function has to take
the form \be Z_{G} = \int_{\Sigma} { {[g^{-1}dg, d{\tilde
A}^l_{z}, d{\tilde A}^l_{\bar z}]} \over {(\textrm{gauge
volume})}} \ \textrm{exp} \left(i S_{G}(g, A_z, A_{\bar z},
J^+, \bar J^+) \right). \ee One must now fix this gauge invariance
to eliminate the non-unique degrees of freedom. One can do this by
employing the BRST formalism which requires the introduction of
Faddev-Popov ghost fields. In order to obtain the
$\it{holomorphic}$ BRST transformations of the fields, one simply
replaces the infinitesimal position-dependent parameters
${\epsilon}^l$ of $h = \textrm{exp}( -
\sum_{l=1}^{\textrm{dim}{\frak n}_+} \epsilon^l t^+_m) \in B$ in the
corresponding $\textrm{\it left-sector}$ of the gauge
transformations in (\ref{gauge tx}) with the ghost fields $c^l$,
which then gives us \be \delta_{\textrm{BRST}}(g) = -c^l t^+_l g,
\quad \delta_{\textrm{BRST}}(\tilde A^l_{\bar z}) = - D_{\bar z}
c^l, \quad \delta_{\textrm{BRST}}(\textrm{others}) =0. \label{BRST
tx SL(N)} \ee The components of the ghost field $c(z) =
\sum_{l=1}^{\textrm{dim}{\frak n}_+} c^l (z) t^+_l$ and those of
its anti-ghost partner $b(z) = \sum_{l=1}^{\textrm{dim}{\frak
n}_+} b^l (z) t^+_l$ will transform as \be \delta_{\textrm{BRST}}
(c^l) = - {1\over 2}f_{mk}^l c^mc^k, \quad
\delta_{\textrm{BRST}}(b^l) = {\tilde B}^l, \quad
\delta_{\textrm{BRST}} {(\tilde B^l)} = 0, \ee where the
$f^l_{mk}$'s are the structure constants of the nilpotent
subalgebra ${\frak n}_+$. Also, the ${\tilde B}^l$'s are the
Nakanishi-Lautrup auxiliary fields that are the BRST transforms of
the $b^l$'s. They also serve as Lagrange multipliers to impose the
gauge-fixing conditions.

In order to obtain the $\it{antiholomorphic}$ BRST transformations
of the fields, one employs the same recipe with the corresponding
$\textrm{\it right-sector}$ of the gauge transformations in
(\ref{gauge tx}), with the infinitesimal position-dependent gauge
parameter now replaced by the ghost field $\bar c^l$, which then
gives us \be \bar \delta_{\textrm{BRST}}(g) = \bar c^l t^+_l g,
\quad \bar \delta_{\textrm{BRST}}({\tilde A}^l_{z}) = - D_{z} \bar
c^l, \quad \bar\delta_{\textrm{BRST}}(\textrm{others})
=0.\label{BRST tx 1 SL(N)}\ee The components of the ghost field
${\bar c}(\bar z) = \sum_{l=1}^{\textrm{dim}{\frak n}_+} {\bar
c}^l (\bar z) t^+_l$ and those of its anti-ghost partner $\bar
b(\bar z) = \sum_{l=1}^{\textrm{dim}{\frak n}_+} {\bar b}^l (\bar
z) t^+_l$ will transform as \be \bar\delta_{\textrm{BRST}} (\bar
c^l) = - {1\over 2}f_{mk}^l {\bar c}^m {\bar c}^k, \quad
\bar\delta_{\textrm{BRST}}(\bar b^l) = {\tilde {\bar B}^l}, \quad
\bar\delta_{\textrm{BRST}} {(\tilde {\bar B}^l)} = 0. \ee In the
above, the $\tilde {\bar B}^l$'s are the Nakanishi-Lautrup
auxiliary fields that are the antiholomorphic BRST transforms of
the $\bar b^l$ fields. They also serve as Lagrange multipliers to
impose the gauge-fixing conditions.

Since the BRST transformations in (\ref{BRST tx SL(N)}) and
(\ref{BRST tx 1 SL(N)}) are just infinitesimal versions of the
gauge transformations in (\ref{gauge tx}), $S_{{\textrm{B-gauged}}} (g, A_z,
A_{\bar z}, J^+, \bar J^+)$ will be invariant under them. An important point to note is that in addition to $(\delta_{\textrm{BRST}} + \bar\delta_{\textrm{BRST}}) \cdot
(\delta_{\textrm{BRST}} + \bar\delta_{\textrm{BRST}}) = 0$, the
holomorphic and antiholomorphic BRST-variations are also
separately nilpotent, i.e., $\delta^2_{\textrm{BRST}} = 0$ and
$\bar\delta^2_{\textrm{BRST}}=0$. Moreover,
$\delta_{\textrm{BRST}}\cdot \bar\delta_{\textrm{BRST}}= - \bar
\delta_{\textrm{BRST}} \cdot \delta_{\textrm{BRST}}$. This means
that the BRST-cohomology of the $B$-gauged WZW model on $G$
can be decomposed into $\it{independent}$ holomorphic and
antiholomorphic sectors that are just complex conjugate of each
other, and that it can be computed via a spectral sequence,
whereby the first two complexes will be furnished by its
holomorphic and antiholomorphic BRST-cohomologies respectively.
Since we will only be interested in the holomorphic chiral algebra
of the $B$-gauged WZW model on $G$ (which, by the way, is
just identical to its antiholomorphic chiral algebra by a complex
conjugation as mentioned), we shall henceforth focus on the $\it{holomorphic}$
BRST-cohomology of the $B$-gauged WZW model on $G$.

By the usual recipe of the BRST formalism, one can fix the gauge
by adding to the BRST-invariant action $S_{\textrm{B-gauged}} (g, A_z,
A_{\bar z}, J^+, \bar J^+)$, a BRST-exact term. Since the BRST
transformation by $(\delta_{\textrm{BRST}} + \bar
\delta_{\textrm{BRST}})$  is nilpotent, the new total action will
still be BRST-invariant as required. The choice of the BRST-exact
operator will then define the gauge-fixing conditions. A
consistent choice of the BRST-exact operator that will give us the
requisite action for the ghost and anti-ghost fields is \be
S_{\textrm{B-gauged}} (g, A_z, A_{\bar z}, J^+, \bar J^+) +
(\delta_{\textrm{BRST}} + \bar \delta_{\textrm{BRST}})
\left({k'\over 2\pi} \int_{\Sigma} d^2 z \
\sum_{l=1}^{\textrm{dim}{\frak n}_+} {\tilde A}^l_{\bar z} b^l +
{\tilde A}^l_{z} {\bar b}^l \right),\nonumber \ee where one will
indeed have the desired total action, which can be written as
\begin{eqnarray} \label{total desired action for SL(N)}
S_{\textrm{WZW}}(g)  -  {k' \over {2\pi}} \int_{\Sigma} d^2z \ \{
\sum_{l=1}^{\textrm{dim}{\frak n}_+} \left[ {\tilde A}^l_{\bar z}(
J^l_+(z) + {M}^l_+ - \tilde B^l) - {\tilde A}^l_{z}( \bar
J^l_+(\bar z) + {\bar M}^l_+ + \tilde {\bar B}^l)
\right]  \nonumber \\
 - \textrm {Tr} [ {A}_z g {A}_{\bar z} g^{-1} - {A}_z {A}_{\bar
z}] \} +  {k'\over {2\pi}}\int_{\Sigma} d^2z \
\sum_{l=1}^{\textrm{dim}{\frak n}_+} \left ( c^l
D_{\bar z} b^l + + \bar c^l D_{z} \bar b^l \right). \nonumber \\
\end{eqnarray} From the equations of motion by varying the ${\tilde
B}^l$'s, we have the conditions $\tilde A^l_{\bar z} = 0$ for
$l=1,\dots, {\textrm{dim}{\frak n}_+}$. From the equations of
motion by varying the $\tilde {\bar B}^l$'s, we also have the
conditions $\tilde A^l_{z} = 0$ for $l=1,\dots,{\textrm{dim}{\frak
n}_+}$. Thus, the partition function of the $B$-gauged WZW model
can also be expressed as
 \be
Z_{G} = \int [ g^{-1} dg, db, dc, d\bar b, d\bar c] \
\textrm{exp} \left ( iS_{\textrm{WZW}}(g) + {i k'\over
{2\pi}}\int_{\Sigma} d^2z \ \textrm{Tr} (c \cdot \partial_{\bar z}
b) (z) + \textrm{Tr} (\bar c \cdot \partial_{z} \bar b) (\bar z)
\right), \label{Z_SL(N)} \ee where the $\it{holomorphic}$ BRST
variations of the fields which leave the effective action in
(\ref{Z_SL(N)}) $\it{invariant}$ are now given by \begin{eqnarray}
\delta_{\textrm{BRST}} (g) = -c^mt^+_m g, & \quad
\delta_{\textrm{BRST}} (c^l) = -{1\over 2} f^l_{mk}c^m c^k,
\quad \delta_{\textrm{BRST}}(b^l) =  J^l_+ + M^l_+ - f^l_{mk}b^mc^k,\nonumber \\
\hspace{-2.0cm}\quad \delta_{\textrm{BRST}}(\textrm{others}) = 0.
& \label{BRST variations SL(N)}
 \end{eqnarray}

The holomorphic BRST-charge generating the field variations in
(\ref{BRST variations SL(N)}) will be given by \be
Q_{\textrm{BRST}} = \oint {dz \over {2 \pi i}} \left(
\sum_{l=1}^{\textrm{dim}{\frak n}_+} \ c^l (z) (J^l_+(z) + M^l_+)
-{1\over 2} \sum_{l=1}^{\textrm{dim}{\frak n}_+} f^l_{mk}b^mc^lc^k
(z) \right). \label{Q_BRST,WZW SL(N)} \ee The free-field action of
the left-moving ghost fields in (\ref{Z_SL(N)}) implies that we
have the usual OPE's of ($\textrm{dim}{\frak n}_+$) free $bc$
systems. From these free $bc$ OPE's, one can verify that
$Q_{\textrm{BRST}}$ in (\ref{Q_BRST,WZW SL(N)}) will indeed
generate the field variations in (\ref{BRST variations SL(N)}).

Though we did not make this obvious in our discussion
above, by integrating out the $\tilde A^l_{\bar z}$'s in
(\ref{B-gauged WZW action G}), and using
the above conditions $\tilde A^l_{z} =0$ for $l=1,\dots,
\textrm{dim}{\frak n}_+$, we find that we actually have the
relations $(J^l_+(z) + M^l_+) =0$ for $l=1,\dots,
\textrm{dim}{\frak n}_+$. These relations---involving the current
associated to the Borel subalgebra $\frak b$ of the group $B$ that
we are modding out from $G$---will lead us directly to the correct form
of the holomorphic stress tensor for the gauged WZW model without
reference to a coset formalism, as we shall see shortly.

Notice that physically consistent with the holomorphic chiral algebra of the purely
bosonic sector of the twisted sigma-model on $G/B$, there are
currents $J^a(z)$ (given by the $J^l_+(z)$'s, $J^l_-(z)$'s and the $J^l_c(z)'s$) in the holomorphic BRST-cohomology of the non-dynamically $B$-gauged WZW model on $G$, where $a=1,2,
\dots, \textrm{dim}{\frak{sl}}_N$, that generate an affine $G$
OPE-algebra at level $k'$.\footnote{The reason that one has level $k'$ instead of $k$ is because the $\psi^{\bar j}$-independent sector of the holomorphic chiral algebra of the twisted sigma-model on $X=G/B$ is, as explained earlier, described by that of the $B$-gauged WZW model on $G$ up to a ${\overline Q}_+$-exact term involving $b_{i\bar j}$; the fluxes associated with $b_{i \bar j}$ will serve to deform the level, as briefly mentioned in footnote 1.} As such, one can construct a
holomorphic stress tensor using the Sugawara formalism as \be
T_{G}(z) = {: d_{ab}(J^a J^b) (z):\over {k'+ h^{\vee}}}. \ee
However, as shown above, one will have the conditions $J^l_+ =
-M^l_+$ for $l=1,2,\dots, \textrm{dim}{\frak n}_+$. In order that
the conformal dimensions of the $J^l_+$'s be compatible with these
conditions, one must define a modified holomorphic stress tensor:
\be T_{\textrm{modified}}(z) = T_{G}(z) + \vec{l} \cdot
\partial\vec{J_c}(z),\ee where $\vec{J_c}(z)$ is a
$\textrm{rank}(\frak g)$-dimensional vector with components being the $J^l_c$
currents associated to the Cartan subalgebra $\frak c$, and
$\vec{l}$ is a sum of simple, positive roots of $\frak g$. In
order for the above conditions on the $J^l_+$'s to be compatible with
the fact that $Q_{\textrm{BRST}}$ generating the holomorphic
variations of the fields must be a scalar of dimension zero, the
$(\textrm{dim}{\frak n}_+)$-set of left-moving ghost systems
$(b^l, c^l)$ must have conformal dimensions $(h^l, 1-h^l)$ for
$l=1,2,\dots, \textrm{dim}{\frak n}_+$, where $h^l$ is the conformal
dimension of the corresponding $J^l_+$ current  under
$T_{\textrm{modified}}(z)$. With all these points in mind, and by
including the holomorphic stress tensor contribution from the action of the free left-moving ghost fields, we can write the total
holomorphic stress tensor of the $B$-gauged WZW model on $G$
as \be T_{\textrm{B-gauged}}(z)  = {: d_{ab}(J^a J^b) (z):\over {k'+
h^{\vee}}} + \sum_{a=1}^{\textrm{dim}{\frak c}}
\partial_z J^a_c (z) -  \sum_{l
\in \vartriangle_+} \left [ h^l b^{l}\partial_z c^{l}(z)
+ (h^l - 1)(\partial_z b^{l} c^{l})(z) \right],\label{T_total}\ee where
$\vartriangle_+$ is the set of positive roots of $\frak g$,
and $\rho^{\vee}$ is the ``dual Weyl vector'' of $\frak g$, such that for $\alpha \in \vartriangle_+$, we have $(\rho^{\vee}, \alpha) =1$ if and only if $\alpha$ is a simple root of $\frak g$.

\newsubsection{The $B$-Gauged WZW Model on $G$ and the ${\cal W}_{k'}(\widehat{\frak g})$ Algebra}

We shall now show that as one would expected from its role as an equivalent description of the holomorphic chiral algebra of the purely bosonic sector of the twisted sigma-model on $G/B$, the holomorphic BRST-cohomology of the $B$-gauged WZW model on $G$ will contain local operators whose Laurent modes generate a ${\cal W}_{k'}(\widehat{\frak g})$ algebra.

To this end, let us first review a purely algebraic approach to generating
${\cal W}_{k'}(\widehat{\frak g})$, the $\cal W$-algebra
associated to the affine algebra $\widehat{\frak g}$
at level $k'$. This approach is known as the quantum
Drinfeld-Sokolov (DS) reduction scheme \cite{Frenkel,
DS-reduction}.

In general, the quantum DS-reduction scheme can be summarised as
the following steps. Firstly, one starts with a triple
$(\widehat{\frak g}, \widehat{\frak g}', \chi)$, where
$\widehat{\frak g}'$ is an affine subalgebra of $\widehat{\frak
g}$ at level $k'$, and $\chi$ is a 1-dimensional representation of
$\widehat{\frak g}'$. Next, one imposes the first class
constraints $ g \sim \chi(g)$ , $\forall g \in {\widehat{\frak
g}'}$, via a BRST procedure. The cohomology of the BRST operator
$Q$ on the set of normal-ordered expressions in currents, ghosts
and their derivatives, is what is called the Hecke algebra
$H^i_Q(\widehat{\frak g}, \widehat{\frak g}', \chi)$ of the triple
$(\widehat{\frak g}, \widehat{\frak g}', \chi)$. For generic
values of $k'$, the Hecke algebra vanishes for $i \neq 0$, and the
existing zeroth cohomology $H^0_Q (\widehat{\frak g},
\widehat{\frak g}', \chi)$, is just spanned by a set of local
operators associated to the triple $(\widehat{\frak g},
\widehat{\frak g}', \chi)$, whose Laurent modes generate a closed
$\cal W$-algebra. We shall denote the $\cal W$-algebra associated
with this set of operators  as ${\cal W}_{DS}[\widehat{\frak g},
\widehat{\frak g}', \chi]$. Note that ${\cal
W}_{DS}[\widehat{\frak g}, \widehat{\frak g}', \chi]$ is
just ${\cal W}_{k'}(\widehat{\frak g})$. Let us
be more explicit about how one can go about defining ${\cal
W}_{DS}[\widehat{\frak g}, \widehat {\frak g}', \chi]$ and
therefore ${\cal W}_{k'}(\widehat{\frak g})$,  now that we
have sketched the general idea behind the DS-reduction scheme.

In order for ${\cal W}_{DS} [\widehat{\frak g}, \widehat{\frak
g}', \chi]$ to be a ${\cal W}$-algebra, one has to suitably choose
the triple $(\widehat{\frak g}, \widehat{\frak g}',\chi)$. A
suitable triple can be obtained by considering a principal
$\frak{sl}_2$ embedding in $\frak g$. Let us now describe this
embedding. Suppose we have an $\frak {sl}_2$ subalgebra $\{t_3,
t_+, t_-\}$ of $\frak g$. The adjoint representation of $\frak g$
decomposes into $\frak{sl}_2$ representations of spin $j_k$ , $k =
1,\dots, s$, for example. Then, one may write the $\widehat {\frak
g}$ current $J(z) = \sum^{\textrm{dim}{\frak g}}_a J^a (z) t_a$ as
\be J(z) = \sum^s_{k=1} \sum^{j_k}_{m= -j_k}  J^{k,m} (z) t_{k,m}
\ee where $t_{k,m}$ corresponds to the generator of spin $j_k$ and
isospin $m$ under the $\frak{sl}_2$ subalgebra. In particular, we
have the correspondences $t_{1,1} = t_+$, $t_{1,0} = t_3$ , and
$t_{1,-1} = t_-$. The $\frak {sl}_2$ subalgebra ${t_3, t_+, t_-}$
can be characterized by a ``dual Weyl vector'' $\rho^{\vee}$, i.e., as mentioned above, for $\alpha \in \vartriangle_+$, where $\vartriangle_+$ is
the set of positive roots of $\frak g$, we have $(\rho^{\vee},
\alpha) =1$ if and only if $\alpha$ is a simple root of $\frak g$.
The $\frak{sl}_2$ root $\hat \alpha$ is given by $\hat \alpha =
\rho/(\rho, \rho)$, and $t_3 = \rho \cdot {\frak c}$, where $\frak
c$ is the Cartan sublagebra of $\frak g$.

Take $\widehat {\frak g}'$ to be the affine Lie subalgebra
$\widehat {\frak n}_+$ generated by all $J^{k,m}(z), m > 0$.
Denoting the currents corresponding to positive roots $\alpha$ by
$J^{\alpha}(z)$, and choosing $t_{1,1} = \sum_i e^{\alpha_i}$, one
can then impose the condition (which realises the required
first-class constraint $g \sim \chi(g)$) \be
\chi_{DS}(J^{\alpha}(z)) = 1 \ (\textrm{for simple roots} \
\alpha_i, ), \quad \chi(J^{\alpha}(z)) =0 \  (\textrm{otherwise}).
\label{constraint} \ee

Next, we introduce pairs of ghost fields $(b^{\alpha}(z),
c_{\alpha}(z))$, one for every positive root $\alpha \in
\vartriangle_+$. By definition, they obey the OPE $b_{\alpha}(z)
c_{\beta} (z') \sim \delta_{\alpha \beta} / (z-z')$, where the
$\alpha, \beta$ (and $\gamma$) indices run over the basis of
$\frak n_+$. The BRST operator that is consistent with
(\ref{constraint}) will then be given by $Q = Q_0 + Q_1$, where
\be Q_0 = \oint {dz \over {2\pi i}} \ \left (J^{\alpha}(z)
c_{\alpha}(z) - {1\over 2}f^{\alpha \beta}_{\gamma} (b^{\gamma}
c_{\alpha}c_{\beta})(z) \right) \label{Q_0} \ee is the standard
differential associated to $\widehat{\frak n}_+$, $f^{\alpha
\beta}_{\gamma}$ are the structure constants of ${\frak n}_+$, and
\be Q_1 =   -\oint {dz \over {2\pi i}}\ \chi_{DS}(J^{\alpha}(z))
c_{\alpha}(z). \label{Q_1} \ee They satisfy \be Q^2 = Q^2_0 =
Q^2_1 = \{Q_0, Q_1\} = 0. \label{Q relations} \ee The resulting
$Q$-cohomology is just the Hecke algebra $H^0_Q (\widehat{\frak
g}, \widehat{\frak g}', \chi)$, which is spanned by a set of local
operators whose Laurent modes generate ${\cal
W}_{DS}[\widehat{\frak g}, \widehat{\frak g}', \chi] = {\cal
W}_{k'}(\widehat{\frak g})$. Note that (\ref{Q relations}) implies
that one can compute the Hecke algebra via a spectral sequence of
a double complex with differentials being $Q_0$ and $Q_1$
accordingly---this strategy has indeed been employed in \cite{de
Boer} to compute explicitly the generators of the ${\cal W}_2 =
{\cal W}_{k'}(\widehat{\frak {sl}}_2)$ and ${\cal W}_3 = {\cal
W}_{k'}(\widehat{\frak {sl}}_3)$ OPE algebras with central charges
$c =13-6(k'+2) - 6/(k'+2)$ and $c =50 - 24(k'+3) - 24/(k'+3)$
respectively.

The variation of the various fields under the action of $Q$ can
also be computed using the OPE's of the affine algebra
$\widehat{\frak g}$, the OPE's of the ghost fields, and the
explicit forms of $Q_0$ and $Q_1$ in (\ref{Q_0}) and (\ref{Q_1})
above, and they are given by
\begin{eqnarray}
\label{field variations for DS-1}
\delta c_{\alpha}(z) & = & -\frac{1}{2}f^{\beta \gamma}_{\alpha}(
c_{\beta}c_{\gamma}) (z),  \\
\label{field variations for DS-2} \delta b^{\alpha}(z) & = &
J^{\alpha} (z) - \chi_{DS} (J^{\alpha}(z)) - f^{\alpha
\beta}_{\gamma} (b^{\gamma} c_{\beta})(z).
\end{eqnarray}

Note also that ${\cal W}_{DS}[\widehat{\frak g}, \widehat{\frak
g}', \chi]$ and thus ${\cal W}_{k'}(\widehat{\frak g})$, will at
least contain the Virasoro algebra. The explicit form of the
stress tensor whose Laurent modes will generate the Virasoro
algebra is (after omitting the normal-ordering symbol)\be
T_{DS}(z) = {{d_{ab}J^a(z)J^b(z)} \over {(k'+h^{\vee})}} +
\sum^{\textrm{dim}\frak c}_{c=1} \partial_z J^c(z) + \sum_{\alpha
\in \vartriangle_+} ((\rho^{\vee}, \alpha) -1)b^{\alpha}\partial_z
c_{\alpha}(z) + (\rho^{\vee}, \alpha)(\partial_z b^{\alpha}
c_{\alpha})(z), \label{stress tensor of DS-reduction}\ee where the
$J^c(z)$'s are just the affine currents that are valued in the
Cartan subalgebra $\frak c$ of the Lie algebra $\frak g$. Note
that with respect to $T_{DS}(z)$, the conformal dimensions of the
pair $(b^{\alpha}(z), c_{\alpha}(z))$ will be given by
$(1-(\rho^{\vee}, \alpha), (\rho^{\vee}, \alpha))$. The central
charge of this Virasoro subalgebra and therefore that of ${\cal
W}_{k'}(\widehat{\frak g})$, will be given by \be c(k') =
{{k'\textrm{dim} {\frak g}}\over {(k'+ h^{\vee})}} -
12k'|\rho^{\vee}|^2 - 2 \sum_{\alpha \in {\vartriangle}_+} \left(
6(\rho^{\vee}, \alpha)^2 - 6(\rho^{\vee}, \alpha)
+1\right).\label{central charge of DS-reduction}\ee

Notice at this point about the $B$-gauged WZW model on $B$, that for any $J^l_+$ with $h^l\neq 0$, the corresponding $M^l_+$ constant must be set to zero for consistency. This means from our above discussion, that one can identify $M^l_+$ with $-\chi_{DS}(J^l_+(z))$. With this identification, one can see that the field variations in (\ref{BRST variations SL(N)}) agree with the field variations in (\ref{field variations for DS-1}) and (\ref{field variations for DS-2}). In addition, we find that $Q_{\textrm{BRST}}$ in (\ref{Q_BRST,WZW SL(N)}) also coincides with $Q = Q_0 + Q_1$, where $Q_0$ and $Q_1$ are given in (\ref{Q_0}) and (\ref{Q_1}), respectively. Moreover, $T_{\textrm{B-gauged}}(z)$ of (\ref{T_total}) is just $T_{DS}(z)$ of (\ref{stress tensor of DS-reduction}). Hence, we see that the holomorphic BRST-cohomology of the $B$-gauged WZW model on $G$ physically realises, in all generality, the purely algebraic DS-reduction scheme of generating the Hecke algebra.

We can summarise the results in this section as follows. Let us label the local operators of the Hecke algebra as $T^{(s_i)}_{\textrm{B-gauged}}(z)$, where $i=1, 2, \dots, \textrm{rank}(\frak g)$, $s_i = e_i +1$; the $e_i$'s being the exponents of $\frak g$, and $T^{(s_i)}_{\textrm{B-gauged}}(z)$ are higher spin-$s_i$ analogs of $T_{\textrm{B-gauged}}(z)$, where $T^{(2)}_{\textrm{B-gauged}}(z) = T_{\textrm{B-gauged}}(z)$. Then, we find that the holomorphic BRST-cohomology of the $B$-gauged WZW model on $G$, will be spanned by local operators $T^{(s_i)}_{\textrm{B-gauged}}(z)$ whose Laurent modes will generate a ${\cal W}_{k'}(\widehat{\frak g})$ algebra with central charge (\ref{central charge of DS-reduction}). Consequently, the classical limit of ${\cal W}_{k'}(\widehat {\frak g})$, i.e., ${\cal W}_{\infty}(\widehat {\frak g})$, will be given by the Poisson $\cal W$-algebra generated by the Laurent modes of the classical fields which lie in the
classical, holomorphic  BRST-cohomology of the $B$-gauged WZW
model on $G$, that are the $\it{classical}$ counterparts of the local operators $T^{(s_i)}_{\textrm{B-gauged}}(z)$.  We shall discuss this set of classical fields
next, and their role in an isomorphism of classical $\cal
W$-algebras and a level relation which underlie a geometric
Langlands correspondence for $G$.

\newsection{A Geometric Langlands Correspondence for $G$}

In this section, we will use what we have learnt in $\S$2 and $\S$3 about ${\frak z}(\widehat {\frak g})$, ${\cal W}_{k'}(\widehat {\frak g})$ and the dual description afforded by the $B$-gauged WZW model on $G$, to show that an equivalence---at the level of the holomorphic chiral algebra---between a bosonic string on $G/B$ and a $B$-gauged version of itself on $G$, will necessarily imply an isomorphism ${\frak z}(\widehat {\frak g})\simeq {\cal W}_{\infty}(^L\widehat {\frak g})$ of classical $\cal W$-algebras and the relation $(k+h^{\vee})r^{\vee}= (k' + {^Lh}^{\vee})^{-1}$ which underlie a geometric Langlands correspondence for $G$.

\newsubsection{The Corresponding Classical Chiral Algebra of the $B$-Gauged WZW Model on $G$}

Let us start by determining the observables of the $B$-gauged WZW model on $G$ which correspond to the $S^{(s_i)}(z)$ fields of the $\psi^{\bar j}$-independent, purely bosonic sector of the twisted sigma-model on $X=G/B$. Firstly, since the $S^{(s_i)}(z)$'s lie in the classical, holomorphic chiral algebra, the corresponding observables must also lie in the $\it{classical}$, holomorphic BRST-cohomology (or chiral algebra) of the $B$-gauged WZW model on $G$. Secondly, an observable corresponding to $S^{(s_i)}(z)$ must also have spin $s_i$, and moreover, it must also generate the same (classical) symmetry in the gauged WZW model as that generated by $S^{(s_i)}(z)$ in the sigma-model.

Now, recall that the $\it{quantum}$ definition of the $S^{(s_i)}(z)$'s at $k\neq -h^{\vee}$, is given by $S^{(s_i)}(z) = (k+h^{\vee})T^{(s_i)}(z)$. Since the $S^{(s_i)}(z)$'s cease to exist as quantum operators at $k = -h^{\vee}$, this must also be true of the corresponding observables in the gauged WZW model. Recall also that at $k \neq -h^{\vee}$, the (Laurent modes of the) $S^{(s_i)}(z)$'s generate a ${\cal W}_{k} (\widehat {\frak g})$ algebra. Similarly, the (Laurent modes of the) $T^{(s_i)}_{\textrm{B-gauged}}(z)$'s in the holomorphic chiral algebra of the gauged WZW model---each having spin $s_i$---generate a ${\cal W}_{k'} (\widehat {\frak g})$ algebra too. A little thought will then reveal that one can consistently identify $S^{(s_i)}(z)$ with ${\overline T}^{(s_i)}_{\textrm{classical}}(z)$---the classical field counterpart of the local operator ${\overline T}^{(s_i)}_{\textrm{B-gauged}}(z) = (k+h^{\vee})T^{(s_i)}_{\textrm{B-gauged}}(z)$. One can see this as follows. Firstly, notice that as required, ${\overline
T}^{(s_i)}_{\textrm{classical}}(z)$ is a spin-$s_i$ field that lies in the classical, holomorphic
chiral algebra of the gauged WZW model at $k=-h^{\vee}$---at $k=-h^{\vee}$, ${\overline T}^{(s_i)}_{\textrm{B-gauged}}(z)$ will act by zero in its OPE's with any other operator, i.e., it will cease to exist as a quantum operator, and will reduce to a purely classical observable ${\overline T}^{(s_i)}_{\textrm{classical}}(z)$. Secondly, since the shift in $h^{\vee}$ in the factor $(k+h^{\vee})$ is due to a quantum renormanisation effect as explained earlier, it will mean that ${\overline T}^{(s_i)}_{\textrm{classical}}(z) = -h^{\vee}\cdot T^{(s_i)}_{\textrm{classical}}(z)$ at $k=-h^{\vee}$, where $T^{(s_i)}_{\textrm{classical}}(z)$ is the classical counterpart of $T^{(s_i)}_{\textrm{B-gauged}}(z)$. This means that the ${\overline T}^{(s_i)}_{\textrm{classical}}(z)$'s will generate the same classical $\cal W$-symmetries in the gauged WZW model as those generated by the $S^{(s_i)}(z)$'s in the sigma-model.

In summary, one can identify the local $S^{(s_i)}(z)$ fields in the classical, holomorphic chiral algebra of the $\psi^{\bar j}$-independent, purely bosonic sector of the twisted sigma-model on $X=G/B$, with the local fields ${\overline T}^{(s_i)}_{\textrm{classical}}(z)$ in the classical, holomorphic chiral algebra of the $B$-gauged WZW model on $G$.

\newsubsection{An Isomorphism of Classical $\cal W$-Algebras and a Geometric Langlands Correspondence for $G$}

We have seen how, from an equivalence---at the level of the holomorphic chiral algebra---between the purely bosonic sector of the twisted sigma-model on $G/B$ and the $B$-gauged WZW model on $G$, one can identify the $S^{(s_i)}(z)$'s with the ${\overline T}^{(s_i)}_{\textrm{classical}}(z)$'s. This identification will in turn imply that the Laurent modes of the local $S^{(s_i)}(z)$ and ${\overline T}^{(s_i)}_{\textrm{classical}}(z)$ fields ought to generate the same classical $\cal W$-algebra with identical central charges.

%%%%%%%%%%%%%%%%%%%%%%%%%%%%%%%%%%%%%%%%%%%%%%%%%%%%%%%%%

What is the central charge of the classical $\cal W$-algebra generated by the Laurent modes of
the ${\overline T}^{(s_i)}_{\textrm{classical}}(z)$ fields? To ascertain this, first note that the central charge of any (classical) $\cal W$-algebra will be given by the central charge of its (classical) Virasoro subalgebra. Next, note that the the Virasoro modes $\hat{L}{}^{(2)}_n$ of $T^{(2)}_{\textrm{B-gauged}}(z) = \sum_{n} \hat{L}{}^{(2)}_n z^{-n-2}$,
will obey the following commutator relation \be [\hat{L}{}^{(2)}_n, \hat{L}{}^{(2)}_m] = (n-m) \hat{L}{}^{(2)}_{n+m} + {c(k')\over {12}} (n^3 - n) \delta_{n, -m} \label{virasoro
relation for wzw sl(2)} \ee at the quantum level, where $c(k')$ is given in (\ref{central charge of DS-reduction}). Therefore, the commutator relations involving the $\hat{\overline L}{}^{(2)}_n$ Virasoro modes
of ${\overline T}^{(2)}_{\textrm{B-gauged}}(z) = \sum_{n} \hat{\overline L}{}^{(2)}_n z^{-n-2}$, will be given by \be [\hat{\overline L}{}^{(2)}_n, \hat{\overline L}{}^{(2)}_m] = (k+h^{\vee}) \left[ (n-m) \hat{\overline L}{}^{(2)}_{n+m} + {c(k',k)\over{12}}(n^3 - n) \delta_{n, -m}
\right]\label{virasoro relation for wzw sl(2) scaled},\ee where $c(k',k) = c(k')(k+h^{\vee})$. At $k
= -h^{\vee}$, ${\overline T}^{(2)}_{\textrm{B-gauged}}(z)$ will cease to have a
quantum definition, and it will reduce to its classical
field counterpart ${\overline T}^{(2)}_{\textrm{classical}}(z)$. Consequently,
the $k \to -h^{\vee}$ (and $k' \to \infty$) limit of the commutator
relation in (\ref{virasoro relation for wzw sl(2) scaled}), can be
interpreted as its classical limit. Therefore, one can view the
term $(k+h^{\vee})$ in (\ref{virasoro relation for wzw sl(2) scaled}) as
the parameter $i \hbar$, where $\hbar \to 0$ is equivalent to the
classical limit of the commutator relations. Since in a
quantisation procedure, we go from $\{ {\overline L}{}^{(2)}_n, {\overline L}{}^{(2)}_m \}_{P.B.} \to {1\over {i\hbar}} [\hat{\overline L}{}^{(2)}_n, \hat{\overline L}{}^{(2)}_m]$, going in reverse would give us the
classical Poisson bracket relation \be \{{\overline L}{}^{(2)}_n, {\overline L}{}^{(2)}_m \}_{P.B.} =  (n-m) {\overline L}{}^{(2)}_{n+m} + {{c(k',k)_{k \to -h^{\vee}, k' \to \infty}} \over 12}(n^3 - n) \delta_{n, -m}, \label{virasoro
relation for wzw sl(2) scaled poisson} \ee where ${\overline
T}^{(2)}_{\textrm{classical}}(z) = \sum_{n} {\overline L}{}^{(2)}_n z^{-n-2}$. Hence, the well-defined central charge of the classical ${\cal W}_{\infty}(\widehat{\frak g})$ algebra generated by the Laurent modes ${\overline L}{}^{(s_i)}_m$ of the ${\overline T}^{(s_i)}_{\textrm{classical}}(z)$ fields, will be given by ${c(k',k)_{k \to -h^{\vee}, k' \to \infty}}$.

%%%%%%%%%%%%%%%%%%%%%%%%%%%%%%%%%%%%%%%%%%%%%%%%%%%%%%%

On the other hand, recall from our earlier discussion that the Laurent modes of the $S^{(s_i)}(z)$ fields will generate a classical ${\cal W}_{-h^{\vee}}(\widehat{\frak g})$ algebra that contains a classical Virasoro subalgebra of central charge $c = -h^{\vee} \textrm{dim}(\frak {g})$
given by \be \{{S}^{(2)}_n, {S}^{(2)}_m \}_{P.B.}  = (n-m)
{S}^{(2)}_{n+m} - { h^{\vee} \ \textrm{dim}({\frak{g}}) \over
12} \  (n^3-n) \ \delta_{n, -m}.\label{classical algebra 1}\ee
Hence, the well-defined central charge of the classical ${\cal W}_{-h^{\vee}}(\widehat{\frak g})$ algebra generated by the Laurent modes $S^{(s_i)}_m$ of the $S^{(s_i)}(z)$ fields, will be given by $c = - h^{\vee} \textrm{dim}(\frak {g})$. Therefore, since the classical $\cal W$-algebras generated by the $S^{(s_i)}_n$'s and the ${\overline L}{}^{(s_i)}_n$'s ought to be isomorphic with the same central charge, it means that we must have the
relation \be c(k,k')_{k\to -h^{\vee}, k'\to \infty} = -{h^{\vee}
\textrm{dim}(\frak g)}.\ee

Note at this point that one can rewrite $c(k')$ as~\cite{review}
\be
c(k') = l - 12|\alpha_+ \rho + \alpha_-\rho^{\vee}|^2,
\ee
where $l = \textrm{rank}(\frak g)$, $\alpha_+ \alpha_- = 1$, $\alpha_- = - \sqrt{k'+h^{\vee}}$, and $\rho$ is the Weyl vector. Since in our case of a simply-laced Lie algebra $\frak g$, we have $\frak g = {^L\frak g}$, it will also mean that we have $\rho = \rho^{\vee}$. One can then simplify $c(k')$ to \be
c(k') = l - 12|\rho|^2 \left(k'+ h^{\vee} + {1 \over {k'+ h^{\vee}}} -2\right).
\label{c(k')}
\ee
From the Freudenthal-de Vries strange formula~\cite{CFT textbook Fransceco et al.}
\be
{|\rho|^2 \over {2 h^{\vee}}} = {\textrm{dim}(\frak g) \over 24},
\ee
and the expression for $c(k')$ in (\ref{c(k')}), we find that
\be
c(k',k) = (k+ h^{\vee})\left[l + 2h^{\vee} \textrm{dim}(\frak g) - h^{\vee} \textrm{dim}(\frak g)\left(k'+ h^{\vee} + {1\over {k'+ h^{\vee}}}\right) \right].
\ee
In the limit $k \to -h^{\vee}$ and $k' \to \infty$, we find that
\be
c(k',k)_{k \to -h^{\vee}, k' \to \infty} = -h^{\vee} \textrm{dim}(\frak g)
\ee
$\textrm{\it{if and only if}}$
\be
(k + h^{\vee})(k' + h^{\vee}) = 1.
\ee

Finally, recall that ${\cal W}_{-h^{\vee}}(\widehat{\frak g})\simeq \frak z(\widehat{\frak{g}})$, and since ${\frak g} = {^L\frak g}$ for $\frak g$ simply-laced, we will also have $h^{\vee} = {^Lh^{\vee}}$ and $r^{\vee}=1$. Hence, we see that an equivalence---at the level of the holomorphic chiral algebra---between the purely bosonic, $\psi^{\bar j}$-independent sector of the twisted sigma-model on
$G/B$ and the $B$-gauged WZW model on $G$, would imply an
isomorphism of Poisson algebras \be \frak z(\widehat{\frak g})
\simeq {\cal W}_{\infty}(^L{\widehat{\frak g}}),\label{isomorphism
of W-algebras SL(N)}\ee and the level relation \be
(k+h^{\vee})r^{\vee} = {1 \over {(k' +
{^Lh^{\vee}})}}.\label{langlands level duality for sl(N)}\ee Recall
at this point that the purely bosonic, $\psi^{\bar j}$-independent
sector of the twisted sigma-model on $G/B$, can be described,
via (\ref{Sbosonic explicit}), by a bosonic string on $G/B$.
On the other hand, note that since a bosonic string on a group
manifold $G$ can be described as a WZW model on $G$, it will mean
that the $B$-gauged WZW model on $G$ can be interpreted as a
$B$-gauged bosonic string on $G$. Thus, we see that an
equivalence, at the level of the holomorphic chiral algebra,
between a bosonic string on $G/B$ and a $B$-gauged version of
itself on $G$---which, can be viewed as a consequence of the ubiquitous
notion that one can always physically interpret a geometrical
symmetry of the target space as a gauge symmetry in the worldsheet
theory---will imply an isomorphism of classical $\cal W$-algebras
and a level relation that underlie a geometric Langlands
correspondence for any complex, $ADE$-group $G$! Note that the correspondence between
the $k \to -h^{\vee}$ and $k' \to \infty$ limits (within the
context of the above Poisson algebras) is indeed consistent with
the relation (\ref{langlands level duality for sl(N)}). These
limits define a ``classical'' geometric Langlands correspondence.
A ``quantum'' generalisation of the $G$ correspondence can be
defined for other values of $k$ and $k'$ that satisfy the relation
(\ref{langlands level duality for sl(N)}), but with the
isomorphism of (\ref{isomorphism of W-algebras SL(N)}) replaced by
an isomorphism of $\it{quantum}$ $\cal W$-algebras (derived from a
DS-reduction scheme) associated to $\widehat{\frak {g}}$ at
levels $k$ and $k'$ respectively \cite{Frenkel}.

\newsection{The Hecke Eigensheaves and Hecke Operators}

We shall now show, via the isomorphism ${\frak z}(\widehat {\frak g})\simeq {\cal W}_{\infty}(^L\widehat {\frak g})$ of classical $\cal W$-algebras demonstrated in $\S$4 above, how one can derive a
correspondence between flat holomorphic $^LG$-bundles on the
worldsheet $\Sigma$ and Hecke eigensheaves on the moduli space
$\textrm{Bun}_G$ of holomorphic $G$-bundles on $\Sigma$, where $\Sigma$ is a closed Riemann surface of any genus.\footnote{Note that the twisted sigma-model on $X$ has an anomaly given by $c_1(X)c_1(\Sigma)$. Hence, since $c_1(X) \neq 0$ for any complex flag manifold $X$, the model is anomalous unless $c_1(\Sigma) =0$. However, since we are only working locally on $\Sigma$ via a local coordinate $z$, i.e., our arguments do not make any reference to the global geometry of the worldsheet which might contribute to a non-zero value of $c_1(\Sigma)$, we can ignore this anomaly. Thus, we are free to work with the sigma-model on $\it{any}$ $\Sigma$.} In the process, we shall be able to physically interpret
the Hecke eigensheaves and Hecke operators in terms of the correlation
functions of purely bosonic local operators in the holomorphic
chiral algebra of the twisted $(0,2)$ sigma-model on the complex
flag manifold $G/B$.

\newsubsection{Hecke Eigensheaves on $\textrm{Bun}_{G}$ and Flat $^LG$-Bundles on $\Sigma$}

\bigskip\noindent{\it Local Primary Field Operators}

As we will explain shortly, the correlation functions of local
primary field operators can be associated to the sought-after
Hecke eigensheaves. As such, let us begin by describing these
operators in the twisted $(0,2)$ sigma-model on a complex flag
manifold $X=G/B$. By definition, the holomorphic primary field
operators $\Phi^{\lambda}_s(z)$ of any theory with an affine
$G$ OPE-algebra obey \cite{CFT textbook Fransceco et al.} \be
J^a(z)\Phi^{\lambda}_r (z') \sim - \sum_{s}
{{(t^a_{\lambda})_{rs}\
\Phi^{\lambda}_s}(z')\over{z-z'}},\label{primary field OPE's}\ee
where $t^a_{\lambda}$ is a matrix in the $\lambda$ representation
of ${\frak{sl}}_N$, $r,s = 1, \dots, \textrm{dim}|\lambda|$, and
$a=1,\dots, \textrm{dim}(\frak g)$.

Since the $\Phi^{\lambda}_s (z)$'s obey OPE relations with the
quantum operators $J^a(z)$, it will mean that they, like the
$J^a(z)$'s, must exist as $\it{quantum}$ bosonic operators of the
sigma-model on $X$. And moreover, since (\ref{primary field
OPE's}) and the affine $G$ OPE-algebra at the critical level
generated by the $J^a(z)$'s in the $\overline Q_+$-cohomology of
the quantum sigma-model together form a closed OPE algebra, it
will mean that the $\Phi^{\lambda}_s (z)$'s are also local
operators in the $\overline Q_+$-cohomology of the sigma-model on
$X$ at the quantum level. From our $\overline Q_+$-Cech cohomology
dictionary (as explained in appendix~A of \cite{Langlands 1}), this means that the $\Phi^{\lambda}_s (z)$'s will
correspond to classes in $H^0(X, {\cal O}^{ch}_X)$, i.e., the
global sections of the sheaf ${\cal O}^{ch}_X$ of  CDO's on $X$.
Note that this observation is also consistent with (\ref{primary
field OPE's}), since one can generate other global sections of the
sheaf ${\cal O}^{ch}_X$ from the OPE's of existing global
sections.

The fact that these operators can be described by global sections
of the sheaf of CDO's on $X$ means that they reside within the
purely bosonic sector of the holomorphic chiral algebra of the
underlying sigma-model on $X$. As we shall see, this observation
will serve as a platform for a physical interpretation of the
Hecke eigensheaves.

\bigskip\noindent{\it Space of Coinvariants}

Associated to the correlation functions of the above-described
local primary field operators, is the concept of a space of
coinvariants, which, in its interpretation as a sheaf over the
moduli space of holomorphic $G$-bundles on $\Sigma$ that we will
clarify below, is directly related to the Hecke eigensheaves that
we are looking for. Hence, let us now turn our attention to
describing this space of coinvariants.

Notice that if the twisted sigma-model were to be conformal, i.e.,
$[\overline Q_+, T(z)] = 0$ even after quantum corrections, we
would have a CFT operator-state isomorphism, such that any primary
field operator $\Phi^{\lambda}_s (z)$ would correspond to a state
$|\Phi^{\lambda}_s \rangle$ in the highest-weight representation
of $\widehat{\frak g}$. However, since
the twisted sigma-model on a complex flag manifold $G/B$ lacks
a holomorphic stress tensor and is thus non-conformal, a
$\Phi^{\lambda}_s (z)$ operator will not have a one-to-one
correspondence with a state $|\Phi^{\lambda}_s \rangle$. Rather,
the states just furnish a module of the chiral algebra spanned by
the local operators themselves.

Nevertheless, in the axiomatic CFT framework of a theory with an
affine algebra $\widehat {\frak g}$, the operator-state
isomorphism is an axiom that is defined at the outset, and
therefore, any primary field operator will be axiomatically
associated to a state in the highest-weight representation of
$\widehat{\frak g}$. Bearing this in mind, now consider a general
correlation function of $n$ primary field operators such as
$\left< \Phi^{\lambda_1}_s (z_1) \dots \Phi^{\lambda_n}_s
(z_n)\right>$. Note that it can be viewed, in the axiomatic CFT
sense, as a map from a tensor product of $n$ highest-weight
representations of $\widehat{\frak g}$ to a complex number.
Next, consider a variation of the correlation function under a
$\it{global}$ $G$-transformation, i.e., $\delta_{\omega}\left<
\Phi^{\lambda_1}_s (z_1) \dots \Phi^{\lambda_n}_s (z_n)\right> =
\oint_{C} dz \sum_a \omega^a \left< J^a(z) \Phi^{\lambda_1}_s
(z_1) \dots \Phi^{\lambda_n}_s (z_n)\right>$, where $\omega^a$ is
a $\it{position}$-$\it{independent}$ scalar transformation
parameter, and where $C$ is a contour that encircles all the
points $z_1, \dots, z_n$ on $\Sigma$. Since all the $J^a(z)$'s are
dimension-one conserved currents in the $\overline Q_+$-cohomology
of the twisted sigma-model on $G/B$, they will generate a
symmetry of the theory. In other words, we will have
$\delta_{\omega}\left< \Phi^{\lambda_1}_s (z_1) \dots
\Phi^{\lambda_n}_s (z_n)\right> =0$, which is simply a statement
of the global $G$-invariance of any theory with an affine
$G$ algebra. This last statement, together with the one
preceding it, means that a general correlation function of $n$
primary field operators $\left< \Phi^{\lambda_1}_s (z_1) \dots
\Phi^{\lambda_n}_s (z_n)\right>$ will define a ``conformal block''
in the axiomatic CFT sense \cite{Frenkel}. Proceeding from this
mathematical definition of a ``conformal block'', the collection
of operators $\Phi^{\lambda_1}_s (z_1) \dots \Phi^{\lambda_n}_s
(z_n)$ will define a vector $\Phi$ in the dual space of
coinvariants $H_{\frak g}(\Phi^{\lambda_1}_s (z_1) \dots
\Phi^{\lambda_n}_s (z_n))$, whereby the ``conformal block'' or
correlation function $\left< \Phi^{\lambda_1}_s (z_1) \dots
\Phi^{\lambda_n}_s (z_n)\right>$ can be computed as the square
$|\Phi|^2$ of length of $\Phi$ with respect to a hermitian inner
product on $H_{\frak g} (\Phi^{\lambda_1}_s (z_1) \dots
\Phi^{\lambda_n}_s (z_n))$ \cite{Frenkel}. All correlation
functions of primary field operators can be computed once this
inner product is determined.

\vspace{-0.0cm}
\bigskip\noindent{\it Sheaf of Coinvariants on
$\textrm{Bun}_{G}$}

As mentioned above, what will be directly related to the Hecke
eigensheaves is the sheaf of coinvariants on the moduli space
$\textrm{Bun}_G$ of holomorphic $G$-bundles on the worldsheet
$\Sigma$. Let us now describe how this sheaf of coinvariants
arises. However, before we proceed, let us first explain how
holomorphic $G$-bundles on $\Sigma$ can be consistently defined in
the presence of an affine $G$-algebra in the sigma-model on
$X=G/B$.

Recall that for the sigma-model on $X = G/B$, we have the OPE
\be J_a (z) J_b (w) \sim {{k d_{ab}} \over{(z-w)^2}} + \sum_c
f_{ab}{}^c {{J_c(w)}\over {(z-w)}}. \ee where $d_{ab}$ is the
Cartan-Killing metric of $\frak g$. Note also that since the
above dimension-one current operators are holomorphic in $\Sigma$,
they can be expanded in a Laurent expansion around the point $w$
on $\Sigma$ as \be J_a(z) = \sum_n{ {J^n_{a}(w)}
{(z-w)^{-n-1}}}.\ee Consequently, from the above OPE, we will get
the commuator relation \be [J^n_a (w), J^m_b (w)] = \sum_c
f_{ab}{}^c J^{n+m}_c (w) + (k d_{ab})\  n \ \delta_{n+m, 0},\label{ramification 2}\ee
such that the Lie algebra $\frak g$ generated by the
zero-modes of the currents will be given by \be [J^0_a (w), J^0_b
(w)] = \sum_c f_{ab}{}^c J^{0}_c (w).\label{zero-modes}\ee One can then exponentiate
the above generators that span $\frak g$ to define an element
of $G$, and since these generators depend on the point $w$
in $\Sigma$, it will mean that one can, via this exponential map,
consistently define a non-trivial principal $G$-bundle on
$\Sigma$. Moreover, this bundle will be holomorphic as the
underlying generators only vary holomorphically in $w$ on the
worldsheet $\Sigma$.

Let us label the above-described holomorphic $G$-bundle on
$\Sigma$ as $\cal P$.  Then, the space $H_{\frak g}
(\Phi^{\lambda_1}_s (z_1) \dots \Phi^{\lambda_n}_s (z_n))$ of
coinvariants will vary non-trivially under infinitesimal
deformations of $\cal P$. As such, one can define a sheaf of
coinvariants over the space  $\textrm{Bun}_{G}$ of all holomorphic $G$-bundles on
$\Sigma$. Let us justify this
statement next.

Firstly, note that with our description of $\cal P$ via the affine
$G$-algebra of the sigma-model on $X$, there is a mathematical
theorem \cite{book} which states that $\textrm{Bun}_{G}$ is
locally uniformized by the affine $G$-algebra. What this means
is that the tangent space $T_{\cal P} \textrm{Bun}_{G}$ to the
point in $\textrm{Bun}_{G}$ which corresponds to an
$G$-bundle on $\Sigma$ labelled by $\cal P$, will be
isomorphic to the space $H^1(\Sigma, \textrm{End} {\cal P})$
\cite{book}. Moreover, deformations of $\cal P$, which correspond
to displacements from this point in $\textrm{Bun}_{G}$, are
generated by an element $\eta(z) = J^a \eta_a(z)$ of the loop
algebra of $\frak g$, where $\eta_a(z)$ is a
$\it{position}$-$\it{dependent}$ scalar deformation parameter (see
$\S$17.1 of \cite {book} and $\S$7.3 of \cite{Frenkel}). With this
in mind, let us again consider the $n$-point correlation function
$\left< \Phi^{\lambda_1}_s (z_1) \dots \Phi^{\lambda_n}_s (z_n)
\right>$. By inserting $\eta(z)$ into this correlation function,
and computing the contour integral around the points $z_1, \dots,
z_n$, we have $\delta_{\eta} \left < \Phi^{\lambda_1}_s (z_1)
\dots \Phi^{\lambda_n}_s (z_n)\right> = \left< \oint_C dz \
\sum_a \eta_a(z) J^a (z) \Phi^{\lambda_1}_s (z_1) \dots
\Phi^{\lambda_n}_s (z_n) \right>$, where $C$ is a contour which
encircles the points $z_1, \dots, z_n$ on $\Sigma$, and
$\delta_{\eta} \left < \Phi^{\lambda_1}_s (z_1) \dots
\Phi^{\lambda_n}_s (z_n) \right>$ will be the variation of $\left<
\Phi^{\lambda_1}_s (z_1) \dots \Phi^{\lambda_n}_s (z_n) \right>$
under an infinitesimal deformation of $\cal P$ generated by
$\eta(z)$ (see eqn. (7.9) of \cite{Frenkel} and also
\cite{Eguchi}). Note that this variation does not vanish, since
$\eta_a(z)$, unlike $\omega$ earlier, is a position-dependent
parameter of a $\it{local}$ $G$-transformation. Therefore, as
explained above, since the correlation function $\left<
\Phi^{\lambda_1}_s (z_1) \dots \Phi^{\lambda_n}_s (z_n) \right>$
is associated to $\Phi$ in the dual space of coinvariants
$H_{\frak g} (\Phi^{\lambda_1}_s (z_1) \dots
\Phi^{\lambda_n}_s (z_n))$, one can see that $\Phi$ must vary in
$H_{\frak g} (\Phi^{\lambda_1}_s (z_1) \dots
\Phi^{\lambda_n}_s (z_n))$ as one moves infinitesimally along a
path in $\textrm{Bun}_{G}$. Since $\Phi$ is just a vector in
some basis of $H_{\frak g} (\Phi^{\lambda_1}_s (z_1) \dots
\Phi^{\lambda_n}_s (z_n))$, one could instead interpret $\Phi$ to
be fixed, while $H_{\frak g} (\Phi^{\lambda_1}_s (z_1) \dots
\Phi^{\lambda_n}_s (z_n))$ varies as one moves infinitesimally
along a path in $\textrm{Bun}_{G}$, as $\cal P$ is subjected
to infinitesimal deformations. Consequently, we have an
interpretation of a $\textrm {\it sheaf of coinvariants}$ on
$\textrm{Bun}_{G}$, where the fibre of this sheaf over each
point in $\textrm{Bun}_{G}$ is just the space $H_{\frak g}
(\Phi^{\lambda_1}_s (z_1) \dots \Phi^{\lambda_n}_s (z_n))$ of
coinvariants corresponding to a particular bundle $\cal P$ that
one can consistently define over $\Sigma$ using the affine
$G$-algebra of the sigma-model on $X=G/B$. Note howeover,
that since we are dealing with an affine $G$-algebra at the
critical level $k = -h^{\vee}$, the dimension of the space of
coinvariants will vary over different points in
$\textrm{Bun}_{G}$. In other words, the sheaf of coinvariants
on $\textrm{Bun}_{G}$ does not have a structure of a vector
bundle, since the fibre space of a vector bundle must have a fixed
dimension over different points on the base. Put abstractly, this
is because $\widehat{\frak g}$-modules at the critical level
may only be exponentiated to a subgroup of the Kac-Moody group
$\widehat {G}$. Nevertheless, the sheaf of coinvariants is a
twisted $\cal D$-module on $\textrm{Bun}_{G}$ \cite{Frenkel}.

From the above discussion, one can also make the following
physical observation. Notice that the variation $\delta_{\eta}
\left < \Phi^{\lambda_1}_s (z_1) \dots \Phi^{\lambda_n}_s (z_n))
\right> = \left< \oint_C dz \ \sum_a \eta_a(z) J^a (z)
\Phi^{\lambda_1}_s (z_1) \dots \Phi^{\lambda_n}_s (z_n) \right>$
in the correlation function as one moves along
$\textrm{Bun}_{G}$, can be interpreted, at the lowest order in
sigma-model perturbation theory, as a variation in the correlation
function due to a $\it{marginal}$ deformation of the sigma-model
action by the term $\oint dz \ \eta(z)$. Since a deformation of
the action by the dimensionless term $\oint dz \ \eta(z)$ is
tantamount to a displacement in the moduli space of the
sigma-model itself, it will mean that $\delta_{\eta} \left <
\Phi^{\lambda_1}_s (z_1) \dots \Phi^{\lambda_n}_s (z_n)) \right>$
is also the change in the correlation function as one varies the
moduli of the sigma-model. This implies that
$\textrm{Bun}_{G}$ will at least correspond to a subspace of the entire
moduli space of the sigma-model on $X = G/B$. This should come
as no surprise since $\cal P$ is actually associated to the affine
$G$-algebra of the sigma-model on $X = G/B$ as explained, and moreover, the affine $G$-algebra does depend on
the moduli of the sigma-model as mentioned in $\S$2.

 Last but not least, note that the sheaf of coinvariants can also be obtained purely mathematically \cite{Frenkel} via a localisation functor $\Delta$,   which maps the chiral vertex algebra $V_{\chi}$---identifiable with all polynomials $F({\cal J}(z))$ (which exist in the chiral algebra of the twisted
sigma-model on $G/B$) that are defined over the field of
complex numbers and the $c$-number operators $S^{(s_i)}(z)$, and
that are of arbitrary positive degree in the quantum operator
${\cal J}(z) = {1 \over {(-n_1 -1)! \dots (-n_m
-1)!}}:\partial_z^{-n_1 -1}J^{a_1}(z) \dots \partial_z^{-n_m
-1}J^{a_m}(z):$---to the corresponding sheaf $\Delta(V_{\chi})$
of coinvaraints on $\textrm{Bun}_{G}$, where $\chi$ denotes a
parameterisation of $V_{\chi}$ that depends on the choice of the
set of $S^{(s_i)}(z)$ fields for $i=1,\dots, \textrm{rank}(\frak
g)$. In other words, the sheaf of coinvariants will be
parameterised by $\chi$.\footnote{Note that in order to be
consistent with the notation used in the mathematical literature,
we have chosen to use the symbol $\chi$ to label the
parameterisation of $V_{\chi}$. Hopefully, $\chi$ that appears
here and henceforth will not be confused with the one-dimensional
representation $\chi$ of $\widehat{\frak g}'$ in $\S$3.} This
observation is pivotal in the mathematical description of the
correspondence between Hecke eigensheaves on
$\textrm{Bun}_{G}$ and flat holomorphic $^LG$-bundles on
$\Sigma$, via the algebraic CFT approach to the geometric Langlands
program \cite{Frenkel}. As we will explain below, this
parameterisation of the sheaf of coinvariants on
$\textrm{Bun}_{G}$ by the set of $S^{s_i}(z)$ fields can be
shown to arise physically in the sigma-model as well.

\bigskip\noindent{\it A $\frak{z}(\widehat{\frak g})$-Dependent Realisation of the
Affine $G$-Algebra at the Critical Level}

Before one can understand how, within the context of the
sigma-model on $X = G/B$,  the sheaf of coinvariants can be
parameterised by a choice of the set of $S^{s_i}(z)$ fields for $i
= 1,\dots, \textrm{rank}(\frak g)$, it will be necessary for
us to understand how one can achieve a $\frak{z}(\widehat{\frak g})$-dependent realisation of the affine $G$ OPE algebra at $k=-h^{\vee}$ spanned by the set of $J^a(z)$ currents that
correspond to classes in $H^0(X,{\cal O}^{ch}_X)$.

To this end, let us first consider the set of local operators composed out of the $n =\textrm{dim}_{\mathbb C}X$
free $\beta_i(z)$ and $\gamma^i(z)$ fields of the $n$ linear $\beta\gamma$
systems associated to the sheaf of CDO's on $X$:
\begin{eqnarray}
J^i_-(z) &= & \beta^{\alpha_i}(z) + \sum_{\varphi \in \Delta_+}
: P^i_\varphi(\gamma^\alpha(z)) \beta^\varphi(z) :, \\
J^k_c(z) &= & - \sum_{\varphi \in \Delta_+} \varphi(h^k) :
\gamma^\varphi(z)
\beta^\varphi(z) :, \\
J^i_+(z) &= & \sum_{\varphi \in \Delta_+} :
Q^i_\varphi(\gamma^\alpha(z)) \beta^\varphi(z) : + c_i \partial_z
\gamma^{\alpha_i}(z),
\end{eqnarray}
where the subscripts $\{\pm,c\}$ denote a Cartan decomposition of
the Lie algebra $\frak g$ under which the $J(z)$ local
operators can be classified, the superscript
$\alpha_i$ denotes the free field that can be identified with the
$i^{th}$ positive root of $\frak g$ where $i = 1, \dots,
n$, $h^k$ is an element of the Cartan subalgebra of $\frak g$ where $k =1,\dots, \textrm{rank}(\frak g)$, $\varphi(h^k)$ is the $k^{th}$
component of the root $\varphi$, the symbol $\Delta_+$ denotes the
set of positive roots of $\frak g$, the $c_i$'s are complex
constants, and lastly, $P^i_\varphi, Q^i_\varphi$ are some
polynomials in the $\gamma^{\alpha}$ free fields.

Theorem 4.3 of \cite{frenkel lectures wakimoto} tells us that the
Laurent modes of the above set of local operators $\{J^i_{\pm},
J^k_c\}$ generate an affine $G$-algebra at the critical level
$k=-h^{\vee}$, i.e., the set $\{J^i_{\pm}, J^k_c\}$ will span an
affine $G$ OPE-algebra at the critical level $k=-h^{\vee}$. Moreover, the fact that the currents $\{J^i_{\pm}, J^k_c\}$ are composed purely out of free $\beta_i$ and $\gamma^i$ fields, and
the fact that there will always be classes in $H^0(X, {\cal
O}^{ch}_X)$ which correspond to operators that generate an affine
$G$ OPE algebra~\cite{MSV}, will together mean that the set of
currents $\{J^i_{\pm}, J^k_c\}$ must correspond (up to $\overline
Q_+$-exact terms at worst) to classes in $H^0(X, {\cal
O}^{ch}_X)$. Equivalently, this means that
the set of local current operators $\{J^i_{\pm}, J^k_c\}$ will be
$\overline Q_+$-closed and hence lie in the holomorphic chiral
algebra of the twisted sigma-model on $X = G/B$.

Next, let us consider a modification $\{J^{i'}_{\pm}, J^{k'}_c\}$ of the
set of currents $\{J^i_{\pm}, J^k_c\}$, where
\begin{eqnarray}
\label{J^{i'}_-} J^{i'}_-(z) &= & \beta^{\alpha_i}(z) +
\sum_{\varphi \in \Delta_+}
: P^i_\varphi(\gamma^\alpha(z)) \beta^\varphi(z) :, \\
\label{J^{k'}_c} J^{k'}_c(z) &= & - \sum_{\varphi \in \Delta_+}
\varphi(h^k) :
\gamma^\varphi(z)\beta^\varphi(z): + b^i(z), \\
\label{J^{i'}_+} J^{i'}_+(z) &= & \sum_{\varphi \in \Delta_+} :
Q^i_\varphi(\gamma^\alpha(z)) \beta^\varphi(z) : + c_i \partial_z
\gamma^{\alpha_i}(z) + b^i(z) \gamma^{\alpha_i}(z),
\end{eqnarray}
and the $b^i(z)$'s are just classical $c$-number functions that
are holomorphic in $z$ and of conformal dimension one---it can be
Laurent expanded as $b^i(z) = \sum_{n \in \mathbb Z} b^i_n
z^{-n-1}$.\footnote{Note that the explicit expression of the $b^i(z)$'s
cannot be arbitrary. It has to be chosen appropriately to ensure
that the Segal-Sugawara tensor and its higher spin analogs given
by the $S^{(s_i)}(z)$'s, can be identified with the space of
$^L\frak g$-opers on the formal disc $D$ in $\Sigma$ as
necessitated by the isomorphism $\frak z(\widehat {\frak g})
\simeq {\cal W}_{\infty}(^L\widehat {\frak g})$ demonstrated earlier. For example,
the expression of $b(z)$ as ${1\over 2}c(z)$ in the $G=SL(2, \bf{C})$ case of \cite{Langlands 1}
ensures that $S'(z) = {1\over 4} c^2(z) - {1\over 2}\partial_z
c(z)$ can be identified with a projective connection on $D$ for
each choice of $c(z)$. However, since the explicit form of the $b^i(z)$'s will not be required for our discussions, we shall not have anything more to say them.} Since the $b^i(z)$'s are classical fields,
they will not participate as interacting (quantum) fields in any of
the OPE's among the quantum operators $\{J^{i'}_+, J^{i'}_-,
J^{k'}_3\}$. Rather, they will just act as a simple multiplication
on the $\gamma^{\alpha_i}(z)$ and $\beta^{\alpha_i}(z)$ fields, or
functions in them thereof. Moreover, this means that the $b^i(z)$'s must
be trivial in the $\overline Q_+$-cohomology of the twisted
sigma-model on $G/B$ at the quantum level, i.e., it can be
expressed as a $\overline Q_+$-exact term $\{\overline Q_+, \dots
\}$ in the $\it{quantum}$ theory. Now, recall that we had the (non
quantum-corrected) geometrical gluing relation $\gamma^{\alpha_i}
= g^{\alpha_i}(\gamma^{\alpha})$, where each $\gamma^{\alpha_i}$
and $g^{\alpha_i}(\gamma^{\alpha})$ is defined in the open set
$U_1$ and $U_2$ respectively of the intersection $U_1 \cap U_2$ in
$X$. This expression means that the $\gamma^{\alpha_i}$'s define
global sections of the sheaf $\widehat{\cal O}^{ch}_X$. From our
$\overline Q_+$-Cech cohomology dictionary, this will mean that
each $\gamma^{\alpha_i}(z)$ must correspond to an operator in the
twisted sigma-model on $X$ that is annihilated by $\overline Q_+$
at the quantum level. This, together with the fact that $b^i(z)$'s
can be expressed as $\{\overline Q_+, \dots\}$, will mean that the
$b^i(z)\gamma^{\alpha_i}(z)$ term in $J^{i'}_+(z)$ of
(\ref{J^{i'}_+}) above, can be written as a $\overline Q_+$-exact
term $\{\overline Q_+, \dots \}$. Likewise, the $b^i(z)$ term in
$J^{k'}_c (z)$ of (\ref{J^{k'}_c}) can also be written as a
$\overline Q_+$-exact term $\{\overline Q_+, \dots \}$.
Consequently, since ${\overline Q}^2_+ =0$ even at the quantum
level, $\{J^{i'}_+, J^{i'}_-, J^{i'}_3\}$ continues to be a set of
quantum operators that are $\overline Q_+$-closed and
non-$\overline Q_+$-exact, i.e., $\{J^{i'}_+, J^{i'}_-, J^{i'}_3\}$ correspond to classes
in $H^0(X, {\widehat{\cal O}^{ch}_X})$. Since the OPE's of
$\overline Q_+$-exact terms such as $b^i(z)\gamma^{\alpha_i}(z)$
and $b^i(z)$ with the other $\overline Q_+$-closed terms
 such as $(\sum_{\varphi \in \Delta_+} : Q^i_\varphi(\gamma^\alpha)
\beta^\varphi : + c_i \partial_z \gamma^{\alpha_i})$,
$(\beta^{\alpha_i} + \sum_{\varphi \in \Delta_+} :
P^i_\varphi(\gamma^\alpha) \beta^\varphi:)$, and $(- \sum_{\varphi
\in \Delta_+} \varphi(h^k) : \gamma^\varphi\beta^\varphi:)$ that
correspond respectively to the set of original operators $J^i_+$,
$J^i_-$, and $J^k_c$, must again result in $\overline Q_+$-exact
terms that are trivial in $\overline Q_+$-cohomology, they can be
discarded in the OPE's involving the set of operators $\{J^{i'}_+,
J^{i'}_-, J^{i'}_3\}$, i.e., despite being expressed differently
from the set of original operators $\{J^i_+, J^i_-, J^k_c\}$, the
set of operators $\{J^{i'}_+, J^{i'}_-, J^{i'}_c\}$ will persist
to generate an affine $G$ OPE-algebra at the critical level
$k=-h^{\vee}$. In other words, via the set of modified operators
$\{J^{i'}_{\pm}, J^{k'}_c\}$ and their corresponding Laurent
modes, we have a different realisation of the affine $G$-algebra at the critical level $k=-h^{\vee}$. This is consistent
with Theorem 4.7 of \cite{frenkel lectures wakimoto}, which states
that the set $\{J^{i'}_{\pm}, J^{k'}_c\}$ of modified operators
will persist to generate an affine $G$ OPE-algebra at the
critical level $k=-h^{\vee}$.

Obviously, from (\ref{J^{i'}_-})-(\ref{J^{i'}_+}), we see that the
above realisation depends on the choice of the $b^i(z)$'s. What
determines the $b^i(z)$'s then? To answer this, let us first
recall that the Segal-Sugawara tensor $S^{(2)'}(z)$ and its higher
spin analogs $S^{(s_i)'}(z)$ associated to the modified
operators $\{J^{i'}_+, J^{i'}_-, J^{k'}_c\} \in \{J^{a'}\}$, can
be expressed as $S^{(s_i)'}(z) = {\tilde d}_{a_1 a_2 \dots
a_{s_i}}:J^{a_1'} J^{a_2'}\dots J^{a_{s_i}'}(z):$ in the quantum
theory. However, recall also that the original Segal-Sugawara
tensor and its higher spin analogs, expressed as $S^{(s_i)}(z) =
{\tilde d}_{a_1 a_2 \dots a_{s_i}}:J^{a_1} J^{a_2}\dots
J^{a_{s_i}}(z):$ in terms of the original operators $\{J^{i}_+,
J^{i}_-, J^{k}_c\} \in \{J^{a}\}$, act by zero in the quantum
theory. This means that the non-vanishing contributions to any of
the $S^{(s_i)'}(z)$'s come only from terms that involve the
additional $b^i(z)$ fields. In fact, it is true that the
$S^{(s_i)'}(z)$'s also act by zero in the quantum theory at
$k=-h^{\vee}$, since they are also defined via a Sugawara-type
construction which results in their quantum definition being
$S^{(s_i)'}(z)= (k+h^{\vee})T^{(s_i)'}(z)$. In other words, the
$S^{(s_i)'}(z)$'s must be classical $c$-number fields of spin
$s_i$ that are holomorphic in $z$. This implies that the
$S^{(s_i)'}(z)$'s will be expressed solely in terms of the
$c$-number $b^i(z)$ fields. An explicit example of this general
statement has previously been discussed in the case of
$G = SL(2, \bf{C})$ in \cite{Langlands 1}---for $G=SL(2, \bf{C})$, we have the identification
$J^{i'}_+ \leftrightarrow J'_+$, $J^{i'}_- \leftrightarrow J'_-$
$J^{k'}_c \leftrightarrow J'_3$, $S^{(2)'}(z) \leftrightarrow
S'(z)$, $b^i(z) \leftrightarrow {1\over 2}c(z)$ and $S^{(2)'}(z) =
{1\over 4} c^2(z) - {1\over 2} \partial_z c(z)$, whereby the
choice of $S^{(2)'}(z)$ determines $c(z)$. Consequently, a choice
of the set of $S^{(s_i)'}(z)$ fields will determine the $b^i(z)$
fields. Lastly, note that the $S^{(s_i)'}(z)$ fields lie in the
classical holomorphic chiral algebra of the purely bosonic sector
of the twisted sigma-model on $X=G/B$, and their Laurent modes
span the centre $\frak z(\widehat{\frak g})$ of the completed
universal enveloping algebra of $\widehat{\frak g}$ at the
critical level $k=-h^{\vee}$. Hence, we effectively have a $\frak
z(\widehat{\frak g})$-dependent realisation of the affine
$G$ (OPE) algebra at the critical level as claimed.

\bigskip\noindent{\it A $\frak{z}(\widehat{\frak g})$-Dependent Parameterisation  of the
Sheaf of Coinvariants on $\textrm{Bun}_{G}$}

Now that we have seen how one can obtain a $\frak z(\widehat{\frak
g})$-dependent realisation of the affine $G$ (OPE)
algebra at the critical level, we can proceed to explain how,
within the context of the sigma-model on $X=G/B$, the sheaf of
coinvariants on $\textrm{Bun}_{G}$ can be parameterised by a
choice of the fields $S^{s_i}(z)$ for $i = 1,\dots,
\textrm{rank}({\frak g})$.

To this end, notice that since the primary field operators
$\Phi^{\lambda}_s(z)$ are defined via the OPE's with the $J^a(z)$
currents of the $\widehat{\frak g}$ algebra at the critical
level in (\ref{primary field OPE's}), a different realisation of
the $J^a(z)$ currents will also result in a different realisation
of the $\Phi^{\lambda}_s(z)$'s. Consequently, we will have a
$\frak z(\widehat{\frak g})$-dependent realisation of the
primary field operators $\Phi^{\lambda}_s(z)$. This amounts to a
$\frak z(\widehat{\frak g})$-dependent realisation of their
$n$-point correlation functions
$\left<\Phi^{\lambda_1}_s(z_1)\dots
\Phi^{\lambda_n}_s(z_n)\right>$. Since the correlation functions
can be associated to a (vector in the) space of coinvariants as
explained earlier, one will consequently have a $\frak
z(\widehat{\frak g})$-dependent realisation of the sheaf of
coinvariants on $\textrm{Bun}_{G}$ as well, i.e., the sheaf of
coinvariants will be parameterised by a choice of the fields
$S^{s_i}(z)$ for $i = 1,\dots, \textrm{rank}({\frak g})$.

\bigskip\noindent{\it A Correspondence Between Hecke Eigensheaves on $\textrm{Bun}_{G}$ and Flat $^LG$-Bundles on $\Sigma$}

Finally, we shall now demonstrate that the above observation about
a $\frak z(\widehat{\frak g})$-dependent realisation of the
sheaf of coinvariants on $\textrm{Bun}_{G}$, and the isomorphism of
Poisson algebras $\frak z(\widehat {\frak g}) \simeq {\cal
W}_{\infty}(^L\widehat{\frak g})$ discussed in $\S$4, will result in a
correspondence between Hecke eigensheaves on
$\textrm{Bun}_{G}$ and flat holomorphic $^LG$-bundles on
the worldsheet $\Sigma$.

Firstly, note that the classsical $\cal W$-algebra ${\cal W}_{\infty}(^L\widehat{\frak g})$ is isomorphic to
$\textrm{Fun}\ \textrm{Op}_{^L{\frak g}}(D^{\times})$, the
algebra of functions on the space of $^L{\frak g}$-opers on
the punctured disc $D^{\times}$ in $\Sigma$, where an
$^L\frak g$-oper on $\Sigma$ is an $n^{th}$ order differential
operator acting from $\Omega^{-(n-1)/2}$ to $\Omega^{(n+1)/2}$
(where $\Omega$ is the canonical line bundle on $\Sigma$) whose
principal symbol is equal to 1 and subprincipal symbol is equal to
0 \cite{Frenkel}. Roughly speaking, it may be viewed as a (flat)
connection on an $^LG$-bundle on $\Sigma$. In turn,
$\textrm{Fun}\ \textrm{Op}_{^L{\frak g}}(D^{\times})$ is
related to the algebra $\textrm{Fun}\ \textrm{Op}_{^L{\frak
g}}(D)$ of functions on the space of $^L{\frak g}$-opers
on the formal disc $D$ in $\Sigma$, via $\textrm{Fun}\
\textrm{Op}_{^L{\frak g}}(D^{\times}) \simeq {\widetilde
U}(\textrm{Fun}\ \textrm{Op}_{^L{\frak g}}(D))$, where
$\widetilde U$ is a functor from the category of vertex algebras
to the category of Poisson algebras \cite{frenkel lectures
wakimoto}. Since we have an isomorphism of Poisson algebras $\frak
z(\widehat {\frak g}) \simeq {\cal
W}_{\infty}(^L\widehat{\frak g})$, it will mean that the
$S^{(s_i)}(z)$'s will correspond to the components of the
(numeric) $^L{\frak g}$-oper on the formal disc $D$ in
$\Sigma$ \cite{Frenkel}. Hence, a choice of the set of
$S^{(s_i)}(z)$ fields will amount to picking up an $^L{\frak
g}$-oper on $D$. Since any $^L{\frak g}$-oper on $D$ can be
extended to a regular $^L{\frak g}$-oper that is defined globally
on $\Sigma$, it will mean that a choice of the set of
$S^{(s_i)}(z)$ fields will determine a unique $^LG$-bundle on
$\Sigma$ (that admits a structure of an oper $\chi$) with a
holomorphic connection.

Secondly, recall that we have a $\frak z(\widehat{\frak
g})$-dependent realisation of the sheaf of coinvariants on
$\textrm{Bun}_{G}$ which depends on the choice of the fields
$S^{s_i}(z)$ for $i = 1,\dots, \textrm{rank}({\frak g})$.
Hence, from the discussion in the previous paragraph, we see that
we have a correspondence between a flat holomorphic
$^LG$-bundle on $\Sigma$ and a sheaf of coinvariants on
$\textrm{Bun}_{G}$.

Lastly, recall that $\Delta(V_{\chi})$---the sheaf of
of coinvariants on $\textrm{Bun}_{G}$---has a structure of a
$\it{twisted}$ $\cal D$-module on $\textrm{Bun}_{G}$. For a general group $G$, the sought-after Hecke eigensheaf \cite{Beilinson} will be given by a $\cal D$-module $\Delta(V_{\chi})\otimes {\Lambda}_\chi^{-1}$ on $\textrm{Bun}_{G}$ with eigenvalue $E_{\chi}$, where ${{\Lambda}_\chi}$ is an invertible sheaf (i.e., a certain line bundle) on $\textrm{Bun}_G$ equipped with a structure of a twisted $\cal D$-module, and $E_{\chi}$ is the unique $^LG$-bundle corresponding to a particular choice of the set of $S^{(s_i)}(z)$ fields. In the case where $G$ is simply-connected, the Hecke eigensheaf will be given \cite{Frenkel} by the $\it{untwisted}$ holonomic $\cal D$-module
$\Delta(V_{\chi})\otimes {K}^{-1/2}$ on $\textrm{Bun}_{G}$ with
eigenvalue $E_{\chi}$, where $K$ is the canonical line bundle on
$\textrm{Bun}_{G}$. In short, since tensoring with the invertible sheaf ${\Lambda}_\chi$ or the canonical line bundle $K$ on $\textrm{Bun}_{G}$ just maps $\Delta(V_{\chi})$ to $\Delta(V_{\chi})\otimes
{\Lambda}_\chi^{-1}$ or $\Delta(V_{\chi})\otimes
{K}^{-1/2}$ in a one-to-one fashion respectively, we find that we have a one-to-one correspondence
between a Hecke eigensheaf on $\textrm{Bun}_{G}$ and a flat
holomorphic $^LG$-bundle on $\Sigma$, where $\Sigma$ is a closed Riemann surface of any genus, i.e., we have a geometric Langlands correspondence for $G$.\footnote{Note that the above-mentioned flat holomorphic
$^LG$-bundles on $\Sigma$ are restricted to those that have a
structure of an $^L\frak g$-oper on $\Sigma$. The space of
connections of any such bundle only form a half-dimensional
subspace in the moduli stack $\textrm{Loc}_{^LG}$ of the space of
$\it{all}$ connections on a particular flat $^LG$-bundle. Thus, our construction establishes the geometric
Langlands correspondence only partially. However, it turns out
that our construction can be generalised to include all flat
$^LG$-bundles on $\Sigma$ by considering in the correlation
functions more general chiral operators that are labelled by
finite-dimensional representations of $\frak g$, which, in
mathematical terms, is equivalent to making manifest the singular
oper structure of any flat $^LG$-bundle on $\Sigma$
\cite{Frenkel}.}

\bigskip\noindent{\it Physical Interpretation of the Hecke Eigensheaves on $\textrm{Bun}_{G}$}

From all of our above results, we see that one can physically
interpret the Hecke eigensheaf as follows. A local section of the
fibre of the Hecke eigensheaf over a point $p$ in
$\textrm{Bun}_{G}$, will determine, for some holomorphic
$G$-bundle on $\Sigma$ that corresponds to the point $p$ in
the moduli space $\textrm{Bun}_{G}$ of all holomorphic
$G$-bundles on $\Sigma$, the value of any $n$-point
correlation function $\left < \Phi^{\lambda_1}_s (z_1) \dots
\Phi^{\lambda_n}_s (z_n) \right>$ of local bosonic operators in
the holomorphic chiral algebra of the twisted $(0,2)$ sigma-model
on $G/B$. And the geometric Langlands correspondence for $G$ just tells us that for every flat, holomorphic
$^LG$-bundle that can be constructed over $\Sigma$, we have a
unique way of characterising how an $n$-point correlation function
of local bosonic primary operators in the holomorphic chiral
algebra of a $\textrm{\it quasi-topological}$ sigma-model with
$\it{no}$ boundaries like the twisted $(0,2)$ sigma-model on
$G/B$, will vary under the $\it{local}$
$G$-transformations generated by the affine $J^a(z)$ currents
on the worldsheet described earlier.

\newsubsection{Hecke Operators and the Correlation Functions of Local Operators}

Consider the quantum operator ${\cal J}(z) = {1 \over {(-n_1 -1)!
\dots (-n_m -1)!}} : \partial_z^{-n_1 -1}J^{a_1}(z) \dots
\partial_z^{-n_m -1}J^{a_m}(z):$. Note that since the $J^a(z)$'s
are $\overline Q_+$-closed and in the $\overline Q_+$-cohomology
or holomorphic chiral algebra of the sigma-model on $G/B$, so
will ${\cal J} (z)$ or polynomials $F({\cal J}(z))$ of arbitrary
positive degree in ${\cal J}(z)$ (modulo  polynomials of arbitrary
positive degree in the $S^{(s_i)}(z)$ operators which necessarily
act by zero and hence vanish in the quantum theory).\footnote{In
order to show this, first note that $\partial_z J^a(z) = [L_{-1},
J^a(z)] $, where $L_{-1} = \oint dz T(z)$. Since $[\overline
Q_+, J^a(z)]=0$ even at the quantum level, it will mean that
$[\overline Q_+, \partial_zJ^a(z)] = [[\overline Q_+, L_{-1}],
J^a(z)] = \oint dz' [ [\overline Q_+, T(z')], J^a(z)]= \oint
dz' [ \partial_{z'}(R_{i \bar j}\partial_{z'} \phi^i \psi^{\bar
j}), J^a(z)] = 0$. One can then repeat this argument and
show that $[\overline Q_+, \partial^m_zJ^a(z)] =0$ for any $m\geq
1$ at the quantum level, always.}

The set of local operators described by $F({\cal J}(z))$ can be
identified with the mathematically defined chiral vertex algebra
$V_{-h^{\vee}}(\frak g)$ associated to $\widehat{\frak g}$ at the
critical level $k=-h^{\vee}$. The action of the Hecke operator on a Hecke
eigensheaf as defined in the axiomatic CFT sense, is equivalent to
an insertion of an operator that lies in the chiral vertex algebra
given by $m$ copies of $V_{-h^{\vee}}(\frak g)$, i.e.,  $\oplus_m
V_{-h^{\vee}}(\frak g)$ \cite{Frenkel}. Such an operator is again
a polynomial operator of the form $F({\cal J}(z))$. In short, the
action of the Hecke operator is equivalent to inserting into the
correlation functions of local primary field operators of the
twisted $(0,2)$ sigma-model on $G/B$, other local operators
that also lie in the holomorphic chiral algebra of the twisted
$(0,2)$ sigma-model on $G/B$, which, as emphasised earlier, is
a $\textrm{\it quasi-topological}$ sigma-model with $\it{no}$
boundaries. This is to be contrasted with the description of the
Hecke operators (and Hecke eigensheaves) in the gauge-theoretic
approach to the geometric Langlands program, where they are
interpreted as 't Hooft line operators (and D-branes) in a
$\it{topological}$ sigma-model $\it{with}$ boundaries. Our results
therefore provide an alternative physical interpretation of these
abstract objects of the geometric Langlands correspondence for
$G$, to that furnished in the gauge-theoretic approach by
Kapustin and Witten in \cite{KW}.

\newsection{The Cases With Tame and Mild Ramifications}

In this section, we shall discuss the cases of tame and mild ramifications in the geometric Langlands correspondence for $G$. We shall explain how, within our context, tamely-ramified, flat $^LG$-bundles on $\Sigma$ will correspond to categories of Hecke eigensheaves on $\textrm{Bun}_{G,\{y_1,\dots, y_k\}}$---the moduli space of holomorphic $G$-bundles on $\Sigma$ with parabolic structures at the points $\{y_1, \dots, y_k\}$ in $\Sigma$. We will do this for mildly-ramified bundles as well. A physical interpretation of these Hecke eigensheaves in terms of the correlation functions of local operators in the holomorphic chiral algebra of the twisted sigma-model on $X = G/B$, will also be furnished.

\newsubsection{Tamely-Ramified $^LG$-bundles on $\Sigma$ and the Category of Hecke Eigensheaves on $\textrm{Bun}_{G,y_1, \dots ,y_k}$}

In the case of tame ramification, the flat connection of the $^LG$-bundle over $\Sigma$ will be modified. Specifically, at a set of points $\{y_1, y_2, \dots, y_k\}$ on $\Sigma$, the connection will have regular singularities, i.e., it will contain a pole of order 1 at each point. In addition, as one traverses around each of these points, the connection will undergo a unipotent monodromy valued in the conjugacy class of $^LG$. For simplicity of argument, let us henceforth consider the case where we only have a single point $y$; the story for multiple points will be analogous. One may then ask the following question: What does this tamely-ramified $^LG$-bundle on $\Sigma$ correspond to in the context of the geometric Langlands correspondence for G?

In order to answer this question, we will first need to revisit the unramified case. Recall that in the unramified case, the sheaf of coinvariants on $\textrm{Bun}_G$ can be obtained purely mathematically as $\Delta_x(V_{\chi_x})$, where $\Delta_x$ is a localisation functor, and where the subscript $x$ is added for convenience to denote that $D$ which appears in the relation $\textrm{Fun}\ \textrm{Op}_{^L{\frak g}}(D^{\times}) \simeq {\widetilde
U}(\textrm{Fun}\ \textrm{Op}_{^L{\frak g}}(D))$, is actually the formal disc at $x \in \Sigma$, such that $\chi_x$ just reflects the restriction of the corresponding $^L{\frak g}$-oper to $D_x$; we omitted this specification earlier as our results in $\S$5 were independent of the point $x$---indeed, we have $\Delta_x(V_{\chi_y}) \simeq \Delta_y(V_{\chi_x})$, where $y$ is any other point in $\Sigma$~\cite{Frenkel}. However, it will be useful to do so for our present discussion on tame ramification.

Note that the chiral vertex algebra $V_{\chi_x}$ is formally called a $(\widehat{\frak g}_x, G_x)$-module because it furnishes a representation of $\widehat{\frak g}_x$, and because the centre $\frak z(\widehat{\frak g}_x)$ commutes with the zero modes of $\widehat{\frak g}_x$ which generate the Lie algebra ${\frak g}_x$ of the group $G_x$. It can be viewed as an object in the category ${\cal C}_{G_x, \chi_x}$ of $(\widehat{\frak g}_x, G_x)$-modules. However, it follows from the results in \cite{frenkel 112} that ${\cal C}_{G_x, \chi_x}$ is simply a category of vector spaces, and its unique up to isomorphism irreducible object is just $V_{\chi_x}$. As such, the localisation functor $\Delta_x$---which actually maps a category of objects to another category of objects---just maps $V_{\chi_x}$ to a unique, irreducible Hecke eigensheaf on $\textrm{Bun}_G$, as discussed in $\S$5.

In the case where the $^LG$-bundle on $\Sigma$ has a tame ramification at say the point $y$, the story will be somewhat different. The relevant oper which describes such a bundle is a $\it{nilpotent}$ $^L{\frak g}$-oper on $D_y$ introduced in \cite{frenkel 44}, and the space $\textrm{Op}^{\textrm{nil}}_{^L{\frak g}}(D_y)$ of such opers is a $\it{subspace}$ of $\textrm{Op}_{^L{\frak g}}(D_y)$. Consequently, we have the relation ${\frak z}({\widehat {\frak g}}_y)\simeq {\widetilde U}(\textrm{Fun}\ \textrm{Op}^{\textrm{nil}}_{^L{\frak g}}(D_y))$, where ${\frak z}({\widehat {\frak g}}_y) \subset {\frak z}({\widehat {\frak g}})$.

 In this ramified case, the object replacing  $V_{\chi_y}$ will be a $(\widehat{\frak g}_y, I_y)$-module, where $I_y$ is an Iwahori subgroup of the loop-group of $G$ that is homomorphic to $B$, the Borel subgroup of $G$ \cite{ramification}; in axiomatic CFT language, the $(\widehat{\frak g}_y, I_y)$-module is a Verma module of $\widehat {\frak g}_y$ at the critical level spanned by vectors which are $I_y$-invariant only. In contrast to the unramified case, the category ${\cal C}_{I_y, \chi_y}$ of $(\widehat{\frak g}_y, I_y)$-modules does not contain a unique irreducible object. Consequently, the localisation functor $\Delta_y$ will map ${\cal C}_{I_y, \chi_y}$ to a $\it{category}$ $\Delta_y({\cal C}_{I_y, \chi_y})$ of Hecke eigensheaves.\footnote{We have, for notational simplicity, omitted the factor $\Lambda^{-1}_{\chi_y}$ that one is supposed to tensor with $\Delta_y({\cal C}_{I_y, \chi_y})$ to get a category of Hecke eigensheaves on the appropriate moduli space to be mentioned briefly.} A Hecke eigensheaf in this category will have an eigenvalue $E_y$, where $E_y$ is a holomorphic $^LG$-bundle over $\Sigma \setminus y$.

One might now ask: on what kind of space is the above category of Hecke eigensheaves defined over? To answer this question, first note that the centre ${\frak z}(\widehat {\frak g}_y)$ commutes with the Lie algebra $\frak b$ of $B$ instead of the Lie algebra $\frak g$ of $G$. Since the centre ${\frak z}(\widehat {\frak g}_y)$ is by definition what commutes with every element of $\widehat {\frak g}_y$, it means that over the point $y$, $\widehat {\frak g}_y$ is effectively $\widehat{\frak b}$, the affine algebra of $B \subset G$; this is consistent with ${\frak z}({\widehat {\frak g}}_y) \subset {\frak z}({\widehat {\frak g}})$. In other words, the commutator relation of (\ref{zero-modes}) will reduce to the commutator relation for the Lie algebra $\frak b$, at $w = y$. Via the exponential map discussed below (\ref{zero-modes}), we see that we actually have a holomorphic  $G$-bundle over $\Sigma$ whose fibre at the point $y$ will be reduced to $B \subset G$---that is, we have a holomorphic  $G$-bundle on $\Sigma$ with $\it{parabolic}$ $\it{structure}$ at the point $y$ in $\Sigma$. Hence, the corresponding category of Hecke eigensheaves will be defined over $\textrm{Bun}_{G, y}$---the moduli space of holomorphic $G$-bundles on $\Sigma$ with parabolic structure at $y$.

If we now consider another point $x$ in $\Sigma$ where there is no ramification of the $^LG$-bundle, the relevant category of modules will be given by ${\cal C}_{G_x, \chi_x}$. However, the category $\Delta_x({\cal C}_{G_x, \chi_x})$ cannot be supported over $\textrm{Bun}_{G,y}$---this is because $\textrm{Bun}_{G,y}$ is an $I_y$-equivariant space, but $\Delta_x({\cal C}_{G_x, \chi_x})$ is not such a category. In other words, the category of $\it{all}$ Hecke eigensheaves on $\textrm{Bun}_{G, y}$ will be given by $\Delta_y({\cal C}_{I_y, \chi_y})$.

Clearly, the above arguments can be easily extended to the multi-point case. In summary, in the geometric Langlands correspondence for $G$ with tame ramification, we have a correspondence between a flat $^LG$-bundle that is tamely-ramified at a set of points $\{y_1, \dots, y_k\}$ on $\Sigma$, and a category of Hecke eigensheaves on $\textrm{Bun}_{G, y_1,\dots, y_k}$---the moduli space of holomorphic $G$-bundles that have parabolic structures at the set of points $\{y_1, \dots, y_k\}$ on $\Sigma$. In addition, a Hecke eigensheaf from the category will have an eigenvalue $E_{y_1, \dots, y_k}$, where $E_{y_1, \dots, y_k}$ is a holomorphic $^LG$-bundle over $\Sigma \setminus \{y_1, \dots, y_k\}$.

\newsubsection{Physical interpretation of Hecke Eigensheaves on $\textrm{Bun}_{G, y_1, \dots, y_k}$}

Recall from our discussion in $\S$5.1, that the variation of an arbitrary correlation function $\left<\Phi^{\lambda_1}_s (z_1) \dots \Phi^{\lambda_n}_s (z_n)\right>$ as one moves infinitesimal in $\textrm{Bun}_G$, will be given by $\delta_{\eta}
\left < \Phi^{\lambda_1}_s (z_1) \dots \Phi^{\lambda_n}_s (z_n)
\right>
\newline
 = \left< \oint_C dz \ \sum_a \eta_a(z) J^a (z)
\Phi^{\lambda_1}_s (z_1) \dots \Phi^{\lambda_n}_s (z_n) \right>$. Also recall that this variation defines the manner in which the corresponding Hecke eigensheaf $\Delta(V_{\chi})$~\footnote{We have, for notational simplicity, omitted the factor $\Lambda^{-1}_{\chi}$ that one is supposed to tensor with $\Delta(V_{\chi})$ to get a Hecke eigensheaf on $\textrm{Bun}_G$.} will vary as one move along $\textrm{Bun}_G$, i.e., it defines a connection on the Hecke eigensheaf $\Delta(V_{\chi})$ over $\textrm{Bun}_G$.

Certainly the connection on a Hecke eigensheaf over $\textrm{Bun}_{G,y}$ will be different as the base space is no longer the same. Consequently, the variation of the correlation function $\left<\Phi^{\lambda_1}_s (z_1) \dots \Phi^{\lambda_n}_s (z_n)\right>$ has to be modified to express this difference. Essentially, one has to insert in the correlation function $\left<\Phi^{\lambda_1}_s (z_1) \dots \Phi^{\lambda_n}_s (z_n)\right>$, a vertex operator which is associated---in the axiomatic CFT sense---to a highest weight vector in the $(\widehat {\frak g}_y, I_y)$-module, at the point $y$ in $\Sigma$~\cite{Frenkel}.

Let us now ascertain what this vertex operator must correspond to in the context of the twisted sigma-model on $X=G/B$. Firstly, note that a $(\widehat {\frak g}_y, I_y)$-module is given by a Verma module in the sense of axiomatic CFT~\cite{Frenkel}. Secondly, recall that the $(\widehat {\frak g}_y, I_y)$-module consists only of $I_y$-invariant vectors. Thirdly, a highest weight vector $\psi$ in a Verma module of an affine algebra $\widehat {\frak g}_y$, is axiomatically defined as a state $|\psi \rangle$, where $J^{\alpha}_n |\psi \rangle =0$ for $n > 0$, and where the $J^{\alpha}_n$'s for $\alpha=1, \dots, \textrm{dim}({\frak b})$ are the generators of $\widehat {\frak g}_y = \widehat{\frak b}$ at $y \in \Sigma$. Altogether, this means that a vertex operator $\varphi(z)$ of our interest, will be axiomatically represented by a state $|\varphi\rangle$, for which $J^{\alpha}_n|\varphi\rangle = 0$ if $n\geq 0$. Notice that such a relation is realised by the OPE
\be
J^{\alpha}(y)\cdot \varphi(w) \sim \textrm{regular},
\label{vertex operator}
\ee
where $y$ is fixed and $w$ is variable in $\Sigma$. Notice that the regular term on the right-hand-side of (\ref{vertex operator}) is a holomorphic function in $w$, and because $\Sigma$ is a compact Riemann surface without boundaries, it will mean that this term is just a constant. Since a constant and the $J^{\alpha}(y)$ currents are invariant under the symmetry transformations generated by the scalar supercharge $\overline Q_+$ of the twisted sigma-model, (\ref{vertex operator}) will imply that $\varphi(w)$ is also $\overline Q_+$-invariant and in the $\overline Q_+$-cohomology. In fact, $\varphi(w)$ corresponds to a class in $H^0(X, {\widehat {\cal O}^{ch}_X})$, i.e., it is a $\psi^{\bar j}$-independent operator in the $\overline Q_+$-cohomology---the cup product of sheaf cohomologies map products of global sections to global sections, and since the $J^{\alpha}(y)$'s correspond to global sections of ${\widehat {\cal O}^{ch}_X}$, and since for $X = G/B$, the space of dimension-zero global sections of ${\widehat {\cal O}^{ch}_X}$ is one-dimensional and spanned by a constant~\cite{MSV}, the OPE (\ref{vertex operator}) will imply that $\varphi(w)$ corresponds to a global section of ${\widehat {\cal O}^{ch}_X}$.

It is readily apparent that the above arguments can be easily extended to the multi-point case.  In summary, the physical interpretation of a Hecke eigensheaf on $\textrm{Bun}_{G, y_1, \dots, y_k}$ in the tamely-ramified case, will be as described in $\S$5---that is, it is (up to a twist by the line bundle $\Lambda^{-1}_{\chi_y}$) the sheaf of coinvariants spanned by  vectors whose lengths-squared give us the values of the corresponding correlation functions of purely bosonic local operators $\Phi^{\lambda_i}_s(z)$ in the holomorphic chiral algebra of the $\it{closed}$, $\it{psuedo}$-$\it{topological}$ twisted sigma-model on $X=G/B$---the only difference being that one has to insert in the correlation functions the local operators $\varphi(y_1), \varphi(y_2), \dots, \varphi(y_k)$ at the ramification points $\{y_1, \dots, y_k\}$ in $\Sigma$ which obey (\ref{vertex operator}), where $\varphi(y_1), \varphi(y_2), \dots, \varphi(y_k)$ are also in the holomorphic chiral algebra of the $\it{closed}$, $\it{psuedo}$-$\it{topological}$ twisted sigma-model on $X=G/B$.

\newsubsection{The Case With Mild Ramification}

Lastly, let us discuss the case with mild ramification. In this case, the flat connection of the $^LG$-bundle over $\Sigma$ will instead have an irregular singularity at each of the points $\{y_1, y_2, \dots, y_k\}$ on $\Sigma$, i.e., it will contain a pole of order $p$, where $1< p \leq n$ for some integer $n$, at each point. Again, for simplicity of illustration, let us consider the situation in which we only have a single point $y$; the story for multiple points will be analogous.

In such a situation, one can just replace the Iwahori subgroup $I_y$ with a congruence subgroup $K_{m,y}$ (with $m \geq n$) in the above arguments of $\S$6.1, and proceed as before~\cite{ramification}. Here, $K_{m,y} = \textrm{exp}\left(\frak g \otimes (\frak m_y)^m\right)$, where ${\frak m}_y$ is the maximal ideal of the ring of integers ${\cal O}_y$ at the point $y$.

In particular, this means that the corresponding category of Hecke eigensheaves will be defined over $\widetilde{\textrm{Bun}}_{G,y}$, the space of holomorphic $G$-bundles whose fibre is reduced to a subgroup of $G$ that is homomorphic to $K_{m,y}$ at the point $y$ on $\Sigma$. In addition, a Hecke eigensheaf in this category has an eigenvalue $\widetilde E$, where $\widetilde E$ corresponds to a flat $^LG$-bundle with mild ramification at the point $y$ on $\Sigma$.

The physical interpretation of a Hecke eigensheaf in this mildly-ramified case will be somewhat similar as before; at the ramification point $y$, one will need to insert in the correlation function $\left<\Phi^{\lambda_1}_s (z_1) \dots \Phi^{\lambda_n}_s (z_n)\right>$, a $\psi^{\bar j}$-independent local operator $\widetilde{\varphi}(y)$ in the holomorphic chiral algebra of the twisted sigma-model on $X=G/B$, that obeys
\be
J^{\widetilde{\alpha}}(y)\cdot \widetilde{\varphi}(w) \sim \textrm{regular},
\label{vertex operator 2}
\ee
where $\widetilde{\alpha} = 1, 2, \dots, \textrm{dim}(\widetilde{\frak g})$;~$\widetilde{\frak g}$ being the Lie algebra of the subgroup of $G$ that is homomorphic to $K_{m,y}$.

The case of mild ramification at multiple points in $\Sigma$ is analogous as one can easily see via a straightforward extension of our above arguments. In summary, the physical interpretation of a Hecke eigensheaf on $\widetilde{\textrm{Bun}}_{G, y_1, \dots, y_k}$ in the mildly-ramified case, will be as described in $\S$5---that is, it is (up to a twist by the line bundle $\Lambda^{-1}_{\chi_y}$) the sheaf of coinvariants spanned by  vectors whose lengths-squared give us the values of the corresponding correlation functions of purely bosonic local operators $\Phi^{\lambda_i}_s(z)$ in the holomorphic chiral algebra of the $\it{closed}$, $\it{psuedo}$-$\it{topological}$  twisted sigma-model on $X=G/B$---the only difference being that one has to insert in the correlation functions the local operators ${\widetilde\varphi}(y_1), {\widetilde \varphi}(y_2), \dots, {\widetilde \varphi}(y_k)$ at the ramification points $\{y_1, \dots, y_k\}$ in $\Sigma$ which obey (\ref{vertex operator 2}), where ${\widetilde\varphi}(y_1), {\widetilde\varphi}(y_2), \dots, {\widetilde\varphi}(y_k)$ are also in the holomorphic chiral algebra of the $\it{closed}$, $\it{psuedo}$-$\it{topological}$ twisted sigma-model on $X=G/B$.

\vspace{1.0cm}
\hspace{-1.0cm}{\large \bf Acknowledgements:}\\
\vspace{-0.5cm}

I would like to thank A. Chervov and L. Rybnikov for stimulating questions which led to the genesis of this paper. I would also like to thank P. Bouwknegt for helpful email correspondences. This work is supported by the Institute for Advanced Study and the NUS-Overseas Postdoctoral Fellowship.

\end{document}